\documentclass[12pt]{article}
\usepackage{hyperref}
\usepackage{feynmp} 
\newcommand{\marrow}[5]{%
    \fmfcmd{style_def marrow#1
    expr p = drawarrow subpath (1/4, 3/4) of p shifted 6 #2 withpen pencircle scaled 0.4;
    label.#3(btex #4 etex, point 0.5 of p shifted 6 #2);
    enddef;}
    \fmf{marrow#1,tension=0}{#5}}

\newcommand{\marrowshort}[5]{%
    \fmfcmd{style_def marrowshort#1
    expr p = drawarrow subpath (1/4, 2/4) of p shifted 6 #2 withpen pencircle scaled 0.4;
    label.#3(btex #4 etex, point 0.4 of p shifted 7 #2);
    enddef;}
    \fmf{marrowshort#1,tension=0}{#5}}

\newcommand{\marrowup}[5]{%
    \fmfcmd{style_def marrowup#1
    expr p = drawarrow subpath (1/4, 3/4) of p shifted 9 #2 withpen pencircle scaled 0.4;
    label.#3(btex #4 etex, point 0.7 of p shifted 15 #2);
    enddef;}
    \fmf{marrowup#1,tension=0}{#5}}

\usepackage{amsmath}
\usepackage{amsfonts}
\usepackage{amsmath}
\usepackage{amsxtra}
\usepackage{amstext}
\usepackage{amssymb}
\usepackage{color}
\usepackage{graphicx}
\numberwithin{equation}{section}
\newcommand{\be}{\begin{equation}}
\newcommand{\ee}{\end{equation}}
\newcommand{\beq}{\begin{equation}}
\newcommand{\eeq}{\end{equation}}
\newcommand{\ber}{\begin{equation}}
\newcommand{\eer}{\end{equation}}
\newcommand{\ba}{\begin{eqnarray}}
\newcommand{\ea}{\end{eqnarray}}
\newcommand{\bea}{\begin{eqnarray}}
\newcommand{\eea}{\end{eqnarray}}
\newcommand{\nn}{\nonumber}
\newcommand{\e}{\epsilon}
\newcommand{\w}{\omega}
\newcommand{\z}{\zeta}
\begin{document}
\baselineskip=15.5pt \pagestyle{plain} \setcounter{page}{1}
%
\begin{titlepage}

\vskip 0.8cm

\begin{center}

{\Large \bf Hadron structure functions at small $x$ from string
theory}

\vskip 1.cm

{\large {{\bf Ezequiel Koile}{\footnote{\tt
koile@fisica.unlp.edu.ar}}, {\bf Nicolas Kovensky}{\footnote{\tt
nico.koven@fisica.unlp.edu.ar}}, {\bf and Martin
Schvellinger}{\footnote{\tt martin@fisica.unlp.edu.ar}}}}

\vskip 1.cm


{\it IFLP-CCT-La Plata, CONICET and Departamento  de F\'{\i}sica,
Universidad Nacional de La Plata.  Calle 49 y 115, C.C. 67, (1900)
La Plata,  Buenos Aires, Argentina.}

\vspace{1.cm}

{\bf Abstract}

\vspace{1.cm}

\end{center}

Deep inelastic scattering of leptons from hadrons at small
values of the Bjorken parameter $x$ is studied from superstring theory.
In particular, we focus on single-flavored scalar and vector mesons
in the large $N$ limit. This is studied in terms of different
holographic dual models with flavor Dp-branes in type IIA and type
IIB superstring theories, in the strong coupling limit of the
corresponding dual gauge theories. We derive the hadronic tensor
and the structure functions for scalar and polarized vector mesons.
In particular, for polarized vector mesons we obtain the eight structure functions
at small values of the Bjorken parameter. The main result is that we
obtain new relations of the Callan-Gross type for several structure
functions. These relations have similarities for all different
Dp-brane models that we consider. This would suggest their universal
character, and therefore, it is possible that they hold for strongly
coupled QCD in the large $N$ limit.

\noindent

\end{titlepage}

\newpage

{\small \tableofcontents}

\newpage

\section{Introduction and general idea}\label{Introduction}

Deep inelastic scattering (DIS) of leptons from hadrons has played a
key role in understanding the hadron structure, by providing
compelling experimental evidence which confirmed predictions from
Quantum Chromodynamics (QCD). In this process an incoming lepton with
four-momentum $k^\mu$, being $k^0 \equiv E$, emits a virtual
photon with four-momentum $q^\mu = k^{\mu} - k'^{\mu}$. This virtual
photon is absorbed by a hadron with four-momentum $P^\mu$. DIS is an
inclusive process, {\it i.e.} while the outgoing lepton four-momentum
$k'^\mu$ (with $k'^0 \equiv E'$) is measured, the final hadronic state
is not. The differential cross section is proportional to the Lorentz
contraction of a leptonic tensor $l^{\mu\nu}$ with a hadronic tensor
$W_{\mu\nu}(P,q)_{h'\,h}$. The leptonic tensor is straightforwardly
calculated from Quantum Electrodynamics, and for a spin-$\frac{1}{2}$
lepton it reads
\begin{equation}\label{DIS3.2}
l^{\mu\nu}=2 \, [k^\mu k'^\nu + k^\nu k'^\mu -\eta^{\mu\nu} (k\cdot
k'+m_l^2) - i \, \epsilon^{\mu\nu\alpha\beta} \, q_\alpha \,
s_{l\beta}] \, ,
\end{equation}
where $s_{l\beta}$ and $m_l$ indicate the leptonic spin and mass,
respectively. On the other hand, the hadronic tensor is expressed in
terms of the commutator of two electromagnetic currents inside the
hadron. Thus, its computation pertains to the domain of QCD at
strong coupling and therefore it cannot be obtained by using
perturbative Quantum Field Theory (pQFT) methods. This is precisely
where the gauge/string duality ideas become useful in this context
\cite{Polchinski:2002jw}.

The definition of the hadronic tensor is given by the following
expression
\begin{equation}\label{DIS4}
W_{\mu\nu}(P,q)_{h'\,h} = i \int d^{4}x \, e^{iq.x} \,
\langle P,h'|[J_{\mu}(x),J_{\nu}(0)]|P,h \rangle \, ,
\end{equation}
where $P^\mu$ and $P_X^\mu$ stand for the hadronic initial and final
momenta, $h$ and $h'$ are the polarizations of the initial and final
hadronic states. In four-dimensional Minkowski spacetime\footnote{We
use the mostly-plus signature for the flat spacetime metric.} we
have the on-shell conditions $M^2=-P^2$ and $M_X^2=-P_X^2$, where we
have written the initial and final hadronic squared masses,
respectively. The hadronic tensor can be rewritten as a sum of a
small set of terms which come from the most general Lorentz-tensor
decomposition of $W_{\mu\nu}(P,q)_{h'\,h}$, satisfying parity
invariance and time reversal symmetry. In this expansion the factors
multiplying each single term are called {\it structure functions}.
From them it is possible to extract the parton distribution
functions, which give the probability that a hadron contains a given
constituent with a given fraction $x$ of its total momentum $P^\mu$.
This number $x$ is the so-called Bjorken parameter defined as
\begin{equation}
x \equiv -\frac{q^2}{2 P \cdot q} \, ,
\end{equation}
and also we define the parameter $t_B$ as
\begin{equation}
t_B \equiv \frac{P^2}{q^2} \, ,
\end{equation}
whose absolute value is very small in the DIS regime.

If the hadrons were composed
by massless partons the probability of finding a parton with a
momentum $x P^\mu$ would be given by the parton distribution
function $f(x, q^2)$. Moreover, if the partons were free the
distribution functions would become independent of $q^2$, leading to
the Bjorken scaling. However, due to the interactions the parton
distribution functions in QCD depend on both $x$ and $q^2$. Notice
that the hadronic structure functions are dimensionless functions
depending on $P^2$, $P \cdot q$ and $q^{2}$, being their functional
dependence recast in terms of $q^2$, $x$ and $t_B$. The physical
ranges for these variables are $0 < x \le 1$ and $t_B \le 0$.

Since there are not known holographic dual models representing
dynamical baryons we focus on spin-zero and polarized spin-one
hadrons, for which there are such models. 
For spin-zero hadrons the most general Lorentz-tensor
decomposition is \cite{Manohar:1992tz,Hoodbhoy:1988am}\footnote{This
expression and Eq.(\ref{DIS16}) for polarized vector mesons differ
from the corresponding ones from
\cite{Manohar:1992tz,Hoodbhoy:1988am} by a few signs. This is due to
the fact that we use of a mostly-plus metric.}
\begin{equation}\label{DIS16sca}
W_{\mu\nu}^{scalar} = F_{1}\bigg(\eta_{\mu\nu}-\frac{q_\mu
q_\nu}{q^2}\bigg)-\frac{F_{2}}{P \cdot
q}\bigg(P_{\mu}+\frac{q_\nu}{2x}\bigg)\bigg(P_{\nu}+\frac{q_\nu}{2x}\bigg)
\, .
\end{equation}
We can just neglect terms proportional to $q_\mu$ and $q_\nu$ since they
vanish upon contraction with the leptonic tensor, since the leptonic
current is conserved.

On the other hand, for polarized spin-one hadrons the most general
form of the hadronic tensor is \cite{Hoodbhoy:1988am}
\begin{eqnarray}\label{DIS16}
W_{\mu\nu}^{vector} &=& F_{1} \, \eta_{\mu\nu}-\frac{F_{2}}{P \cdot
q}P_{\mu}P_{\nu}+b_{1}r_{\mu\nu}-\frac{b_{2}}{6}(s_{\mu\nu}+t_{\mu\nu}+
u_{\mu\nu}) -\frac{b_{3}}{2}(s_{\mu\nu}-u_{\mu\nu})\nonumber\\ & &
-\frac{b_{4}}{2}(s_{\mu\nu}-t_{\mu\nu}) -\frac{i \, g_{1}}{P \cdot
q} \, \epsilon_{\mu\nu\lambda\sigma} \, q^{\lambda} \, s^{\sigma}
-\frac{i \, g_{2}}{(P \cdot q)^{2}} \,
\epsilon_{\mu\nu\lambda\sigma} \, q^{\lambda}
\, (P \cdot q \:s^{\sigma}-s \cdot q \:P^{\sigma})\, , \nonumber\\
\end{eqnarray}
where $F_1$, $F_2$, $g_1$, $g_2$, $b_1$, $b_2$, $b_3$ and $b_4$ are
the eight structure functions for polarized spin-one hadrons. In
the above equation we have dropped terms proportional to $q_\mu$ and
$q_\nu$, since as before they vanish when they are contracted with
$l^{\mu\nu}$. We have also used the following definitions
\begin{eqnarray}
\label{DIS17} && r_{\mu\nu}\equiv\frac{1}{(P \cdot q)^{2}}\bigg(q
\cdot \zeta^{*}\:q \cdot \zeta-\frac{1}{3}(P \cdot
q)^{2}\tilde\kappa \bigg)\eta_{\mu\nu} \, ,
\end{eqnarray}
\begin{eqnarray}
\label{DIS18} && s_{\mu\nu}\equiv\frac{2}{(P \cdot q)^{3}}\bigg(q
\cdot \zeta^{*}\:q \cdot \zeta-\frac{1}{3}(P \cdot
q)^{2}\tilde\kappa \bigg)P_{\mu}P_{\nu} \, ,
\end{eqnarray}
\begin{eqnarray}
\label{DIS19} && t_{\mu\nu}\equiv\frac{1}{2(P \cdot q)^{2}}\bigg(q
\cdot \zeta^{*}\:P_{\mu}\zeta_{\nu}+q \cdot
\zeta^{*}\:P_{\nu}\zeta_{\mu} +q \cdot
\zeta\:P_{\mu}\zeta^{*}_{\nu}+q \cdot \zeta
\:P_{\nu}\zeta^{*}_{\mu}-\frac{4}{3}(P \cdot q) P_{\mu}P_{\nu}\bigg)
\, , \nonumber \\
&&
\end{eqnarray}
\begin{eqnarray}
\label{DIS20} && u_{\mu\nu}\equiv\frac{1}{P \cdot
q}\bigg(\zeta^{*}_{\mu}\zeta_{\nu}+\zeta^{*}_{\nu}\zeta_{\mu}
-\frac{2}{3}M^{2}\eta_{\mu\nu}
-\frac{2}{3}P_{\mu}P_{\nu}\bigg) \, ,
\end{eqnarray}
being $\tilde\kappa= 1 - 4 \, x^2 \, t_B$ while $s^{\sigma}$ is a
four-vector analogous to the spin four-vector in the case of
spin-$\frac{1}{2}$ fields defined as
$s^\sigma\equiv-\frac{i}{M^2}\epsilon^{\sigma\alpha\beta\tau}\zeta^*_\alpha\zeta_\beta
P_\tau$. We can see the dependence
on the initial and final hadronic polarization vectors denoted by
$\zeta_\mu$ and $\zeta^*_\mu$. The transversality condition $P \cdot
\zeta=0$ is satisfied, while the hadronic polarization vectors are
normalized such that $\zeta\cdot\zeta^*=M^{2}$.

One should also notice that DIS amplitudes can be obtained by taking
the imaginary part of the forward Compton scattering amplitudes.
This allows one to consider the tensor\footnote{Current-current correlation
functions in the DIS regime of an ${\cal {N}}=4$ SYM plasma have been considered
in \cite{Hatta:2007cs} and by including $\alpha'^3$-corrections in \cite{Hassanain:2009xw}.
In addition, in the hydrodynamical regime of an ${\cal {N}}=4$ SYM plasma this kind of 
correlation functions is necessary in order to obtain the electrical conductivity
\cite{CaronHuot:2006te}, which has also been studied by including $\alpha'^3$-corrections
from string theory \cite{Hassanain:2011ce,Hassanain:2011fn,Hassanain:2010fv,Hassanain:2012uj}.}
\begin{equation}\label{DIS50}
T_{\mu\nu}(P,q)_{h'\,h} =
i \, \int d^{4}x \, e^{iq.x} \, \langle P, {\cal
{Q}}|{\widehat{T}}(J_{\mu}(x) \, J_{\nu}(0))|P,  {\cal
{Q}}\rangle \, ,
\end{equation}
where $J_\mu$ and $J_\nu$ are the electromagnetic current operators, 
and ${\cal {Q}}$ is the charge of the hadron.
${\widehat{T}}(\widehat{{\cal{O}}}_{1}\widehat{{\cal{O}}}_{2})$
stands for time-ordered product between the operators
$\widehat{{\cal{O}}}_{1}$ and $\widehat{{\cal{O}}}_{2}$, and the
tilde indicates the Fourier transform of the electromagnetic current
operator. The tensor $T_{\mu\nu}\equiv T_{\mu\nu}(P,q)_{h'\,h}$ has
identical symmetry properties as the hadronic tensor, thus having
a Lorentz-tensor structure similar to $W_{\mu\nu}(P,q)_{h'\,h}$. The
optical theorem leads to
\begin{equation}\label{DIS51}
F_{j} = 2 \pi \, \textmd{Im} \widetilde {F_{j}}\, ,
\end{equation}
where $\widetilde{F_j}$ is the $j$-th structure function of the
$T_{\mu\nu}$ tensor, while $F_j$ is the one corresponding to the
$W_{\mu\nu}$ tensor.

As pointed out before, in order to calculate the hadronic tensor one
cannot approach the problem in terms of perturbative QCD, since the
parton distribution functions depend on soft QCD dynamics.
Polchinski and Strassler \cite{Polchinski:2002jw} developed a
proposal to calculate the hadronic tensor for glueballs by using the
gauge/string theory duality. They obtained the structure functions
for glueballs in a deformation of the large $N$ limit of $SU(N)$
${\cal {N}}=4$ SYM theory, which leads to the ${\cal {N}}=1^*$ SYM
theory \cite{Polchinski:2000uf}. Their calculation holds in the
strongly coupled regime of the gauge theory, {\it i.e.} when the 't
Hooft coupling $\lambda=g_{YM}^2 N$ satisfies $1 \ll \lambda \ll N$.
There are four kinematical regimes depending on the values of the
Bjorken parameter. The supergravity regime  
holds when $1/\sqrt\lambda \ll x < 1$ ($x=1$ corresponds
to elastic scattering). The second kinematic regime
holds provided that $\exp{(-\sqrt\lambda)} \ll x \ll
1/\sqrt\lambda$, and in that case the holographic dual description
corresponds to excited strings. In the third regime one
considers exponentially small values of $x$, which corresponds to
the case when the size of the strings are comparable to the AdS$_5$
scale $R$. This is a subtle but interesting parameter region because
the interaction can no longer be considered local. In this region
there is an effect due to the strings growth which is studied in
terms of a diffusion operator. There is a fourth regime 
where $|\ln x| \lambda^{-1/2} > \ln (\Lambda/q)$, being $q$ the
momentum transfer and $\Lambda$ the confining IR scale. In this
case the world-sheet renormalization group can be used to include
the effect of strings growth. It is worth mentioning that the general
picture in the planar limit of the strongly coupled gauge field
theory corresponds to the scattering to a lepton by an entire hadron.
Within the last three regimes the calculations 
have to be done in terms of superstring theory
scattering amplitudes as we will discuss in detail in the following
sections for holographic mesons. Alternatively, the calculation can
be carried out in a different way, in terms of an effective
Lagrangian which contains a four-point interaction vertex. This
effective Lagrangian is derived from the four-point string theory
scattering amplitude which is written in terms of the product of a
kinematic factor and a pre-factor. The kinematic factor can be
straightforwardly derived by extracting the coefficient of the
graviton pole of the $t$-channel. This can be done within the
low-energy limit of string theory. On the other hand, the pre-factor
contains the $\alpha'$-dependence through gamma functions. This
calculation in fact goes beyond the supergravity approximation.

Obtaining the structure functions in the whole physical range of the
Bjorken parameter is very interesting, particularly at small
$x$. From the study of the moments of the structure functions $F_1$
and $F_2$ it can be inferred that the structure functions should
have a component with a narrow peak around $x=0$. Moreover, there is
a more physical argument supporting this behavior
\cite{Kogut:1974ni}, which can be described as follows. First, let
us consider weakly coupled gauge theory where the interactions
produce splitting of partons and, therefore, the structure functions
increase as the Bjorken parameter decreases. As the coupling
constant increases the evolution of the structure functions becomes
more rapid. At strong coupling we cannot use the parton model.
However, one might think that this trend, which leads to an even
more rapid evolution towards small $x$, should still hold. This has
indeed been confirmed for the glueball structure functions by using
a string theory calculation \cite{Polchinski:2002jw}.

Now, the questions are whether or not
it is possible to extract the eight structure functions
from holographic dynamical hadrons, and
what can be said about the $x$-dependence of the structure functions
for small and exponentially small values of the Bjorken parameter.
We can address these questions in the planar limit and at strong coupling.
As commented before, since there are not known holographic dual
models of dynamical baryons, it is then compelling to consider
dynamical scalar and polarized vector
mesons. This is interesting for several reasons. One is to
understand the structure of the holographic dual mesons.
Also, we are interested in looking
for general properties, {\it i.e.} properties which either do not
depend on the particular holographic dual model, or depend on it in
such a way that we can say something about what would happen to the
structure functions of QCD mesons in the large $N$ limit.

In \cite{Koile:2011aa} we began with this research programme for
dynamical holographic scalar and polarized vector mesons with one
flavor, by considering different flavor Dp-brane models in type IIB
and type IIA superstring theories. Then, in \cite{Koile:2013hba} we
generalized these investigations to the case of several flavors,
which is indeed a non-trivial generalization, and also by obtaining
the corresponding next-to-leading order Lagrangians in the $1/N$ and
$N_f/N$ expansions. In those papers we have considered different
holographic dual models leading to different confining gauge
theories. Particularly, we have studied the D3D7-brane model dual to
an ${\cal {N}}=2$ supersymmetric Yang-Mills theory with fundamental
quarks \cite{Kruczenski:2003be}, and also the gauge theories which
are dual to the D4D8$\mathrm{\overline{D8}}$-brane model of Sakai
and Sugimoto \cite{Sakai:2004cn} and the
D4D6$\mathrm{\overline{D6}}$-brane model \cite{Kruczenski:2003uq},
respectively. For all these different confining gauge theories, we
have obtained the corresponding scalar and polarized vector meson
structure functions in the supergravity limit, {\it i.e.} in the
kinematic region where $1/\sqrt\lambda \ll x < 1$. 
A schematic picture of the holographic dual description of the 
forward Compton scattering within this parametric (pure supergravity) 
regime is depicted in figure 1. 

\begin{figure}
\begin{center}
\includegraphics[scale=0.32]{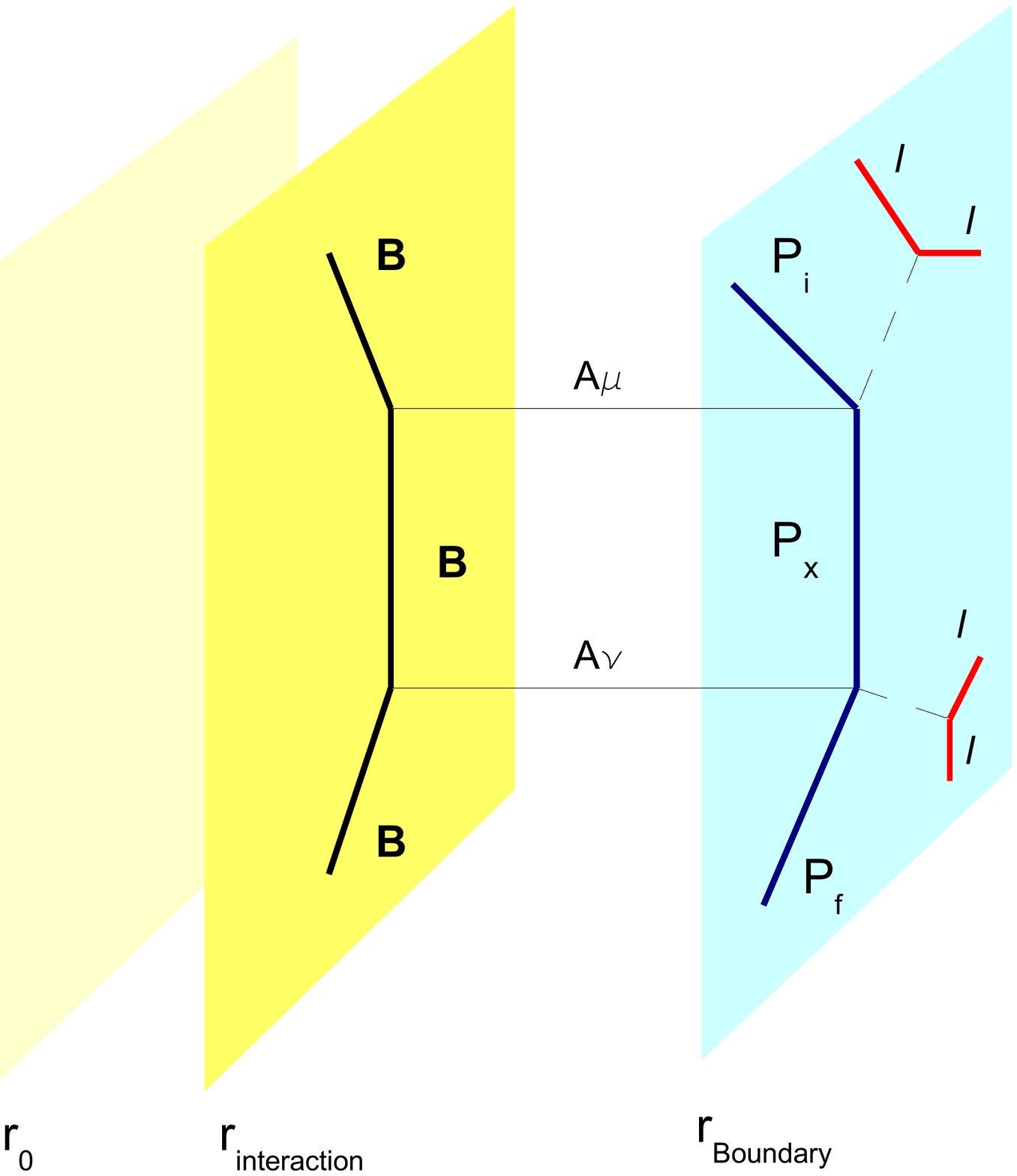}
{\caption{\small A schematic picture of the holographic dual description 
of DIS in the $1/\sqrt\lambda \ll x < 1$ regime. $B$ denotes vector 
mesons, $r$ is the radial coordinate, $r_0$ is the IR cutoff, 
$r_{interaction}$ is where the graviton-meson interaction occurs (the 
position of the D7-brane), and $r_{Boundary}$ is where the actual
field theory DIS process takes place. In this figure $l$ denotes leptons,
dashed lines indicate virtual photons. This is a holographic dual
representation of the forward Compton scattering, related to DIS through 
the optical theorem.}}
\end{center}
\end{figure}

We have found
several interesting results. On the one hand, we have obtained new
relations among several of the eight different structure functions
of each polarized vector meson. On the other hand, we have found
that these relations are independent of the model: there could be a
universal structure for holographic scalar and vector mesons. The
reason for this behavior is the fact that the dynamics of
holographic dual models with probe flavor Dp-branes is controlled by
the Dirac-Born-Infeld (DBI) action. Although for each particular
model the DBI action changes its dimension and also the structure of
the gauge fields and the induced metric, it renders
model-independent Callan-Gross type relations for different Dp-brane
models when $1/\sqrt\lambda \ll x < 1$, for instance $F_2 = 2 F_1$.
Notice though that for small-$x$ the Callan-Gross relation from QCD
has an extra $x$-factor: $F_2 = 2 x F_1$. Thus, it is expected that
this additional factor should be present for small $x$ in the
holographic dual description.

The next question is how to calculate the referred scalar and
polarized vector meson structure functions at small and
exponentially small values of $x$. This is the task we carry out in
the present work. The calculations we perform hold in the planar
limit of the gauge theory. Notice that for simplicity we restrict
ourselves to the case of $N_f=1$, {\it i.e.} single-flavored mesons.

We address three main issues. As mentioned before, one is about the
behavior of the structure functions at small and exponentially small
$x$, in order to see if the rapid evolution towards small values of
$x$ at strong coupling is confirmed for dynamical holographic
mesons. Secondly, we show that for each holographic dual model
we consider one may extract the Callan-Gross relation between $F_1$
and $F_2$, of the form $F_2 \sim 2 x F_1$, and similar additional
relations among other structure functions. In fact as it happens
with the glueballs \cite{Polchinski:2002jw} we find that there is
also an extra $x$-independent (but model dependent) factor on the
right hand side. Thirdly, we show how general these relations are,
{\it i.e.} we discuss on their model-independent behavior. In order
to carry out our programme we have to calculate the structure
functions by using superstring theory. This has to be done in terms
of the scattering amplitudes of two closed and two open strings. We
will show the details in the following sections.

In Section 2 we study DIS of leptons by scalar mesons. Then, in
Section 3 we carry out the calculation of the hadronic tensor of
polarized vector mesons, including the structure functions and their
relations at small values of the Bjorken parameter $x$ from string
theory. We begin with the four-point scattering 
amplitudes of two open and two closed strings, ${\cal {A}}_4^{2o2c,
scalar}$, in flat spacetime. 
This has two terms with two factors each: a kinematic one and
a pre-factor which carries the $\alpha'$-dependence. In the parametric
regime we consider there is only
one relevant term and from its kinematic factor one can obtain an
effective four-point interaction Lagrangian. This Lagrangian
describes an effective interaction between two gravitons and two
scalar mesons. In the case of vector mesons the corresponding string
theory scattering amplitude, ${\cal {A}}_4^{2o2c, vector}$, has
several terms with two factors each: a kinematic one and a
pre-factor. Again, we argue that there is only one relevant term and
explain how one can obtain an effective four-point interaction
Lagrangian. This Lagrangian describes an effective interaction
between two gravitons and two polarized vector mesons.
We should emphasize that this flat spacetime calculation 
is directly related to the scattering process in our curved background.
This is so because in the small $x$ regime the size of the strings is
small compared to the AdS curvature. As we shall see later this implies
that the interaction can be considered local.

From the string theory scattering amplitude ${\cal {A}}_4^{2o2c}$
one can proceed in two different ways. In the first place, one can take the limit
$\tilde {t} \rightarrow 0$, where $\tilde {t}$ is the
ten-dimensional $t$-channel Mandelstam variable. From it one can
obtain an effective Lagrangian, for which the relevant terms within
the kinematic regime we describe consist of several four-point
interaction vertices. Then, from this Lagrangian one can calculate
the hadronic tensor for small and exponentially small values of $x$.
There is a the second approach which we comment as follows. Let us
consider the low-energy action of superstring theory, the Dp-brane
action and the interaction between open and closed superstrings.
Then, let us consider the $\alpha' \rightarrow 0$ limit. From this
low-energy theory we derive the graviton and meson propagators and
also the interaction vertices. Then, we explicitly calculate the
$s$-, $t$- and $u$-channels. It turns out that the coefficient of the
$t$-channel graviton pole gives the same effective four-point
interaction Lagrangian mentioned above up to an $\alpha'$-dependent factor. 
A schematic representation of this process is depicted in figure 2. We carry out this
calculation in full detail, and discuss its connection with the
string theory four-point scattering amplitude.

\begin{figure}
\begin{center}
\includegraphics[scale=0.32]{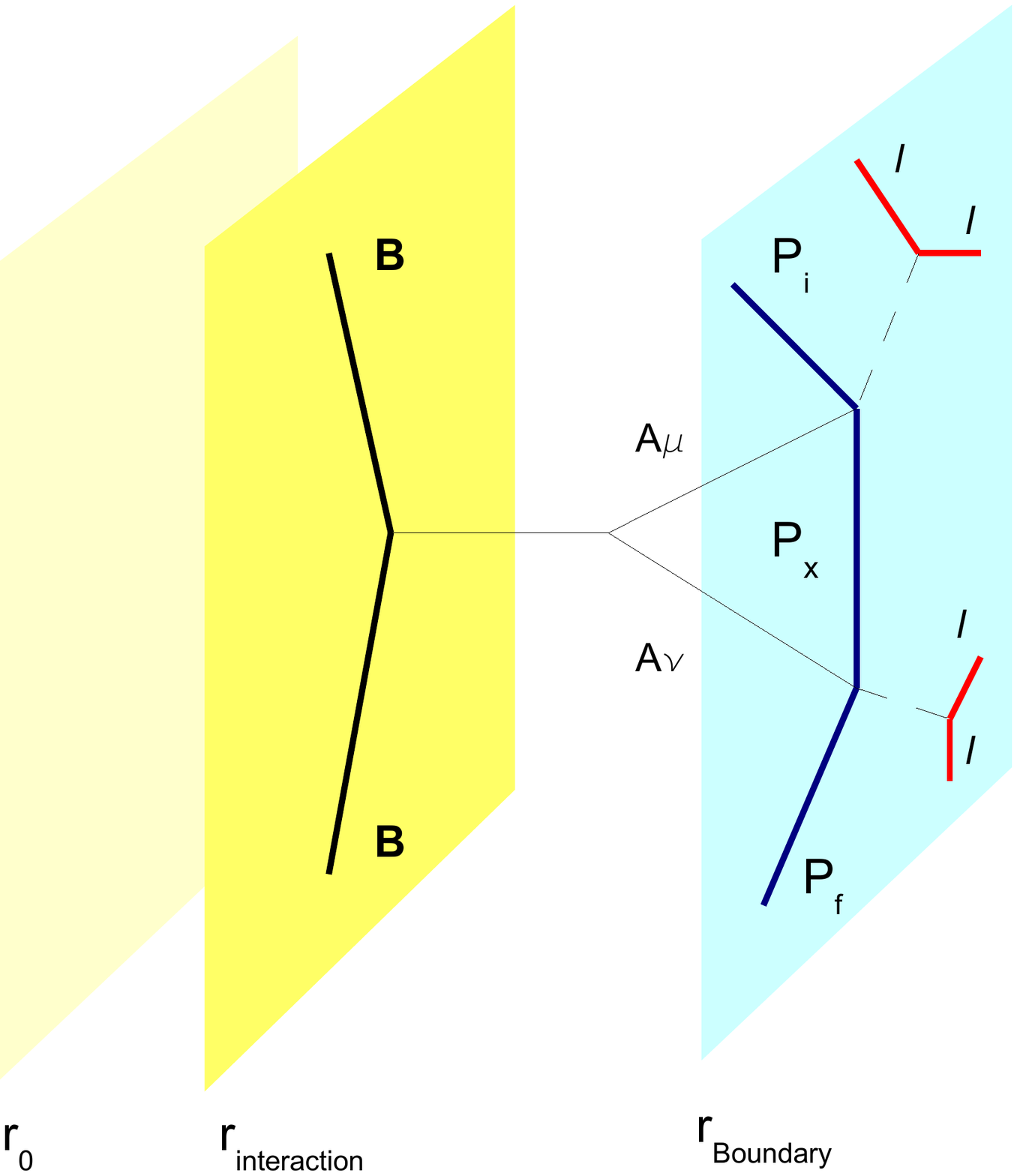}
{\caption{\small Schematic picture of the holographic dual description of DIS
in the $\exp\left(-\sqrt{\lambda}\right)\ll x \ll 1/\sqrt\lambda $ regime.
Here the graviton exchanged between the three-graviton vertex and the 
meson-meson-graviton vertex represents a single Pomeron exchange in the dual
gauge theory.}}
\end{center}
\end{figure}

Then, from these effective Lagrangians we derive the hadronic
tensors of different mesons. Finally, we calculate the structure
functions for scalar and polarized vector mesons. Particularly, for
polarized vector mesons we obtain the alluded eight structure
functions at small and exponentially small $x$ values. Very
interestingly, we obtain new relations of the Callan-Gross
type for several structure functions. These relations have
similarities for all Dp-brane models we consider. This suggests a
universal behavior which would possibly hold in the large $N$ limit
of QCD.

In Sections 2 and 3 we carry out all the mentioned calculations for
the specific case of the D3D7-brane model. We also comment
on the exponentially small $x$ regime. In Section 4 we extend
all these results, expressing them in a compact general form which
also holds for the D4D6$\mathrm{\overline{D6}}$-brane model and for
the D4D8$\mathrm{\overline{D8}}$-brane model. Discussions and
conclusions are presented in Section 5.

\section{DIS from scalar mesons at small $x$ from string theory}

Let us consider the four-momentum of the hadron $P^\mu$,
and the virtual photon four-momentum
$q^\mu$, where $q^2>0$. Recall that the $s$-channel Mandelstam
variable in the four-dimensional gauge theory is given by
\be
s=-(P+q)^2 = q^2 \left( \frac{1}{x} - 1 - t_B \right)
\simeq q^2 \left( \frac{1}{x} - 1 \right) \, ,
\ee
where we have considered the situation where $|P^2| \ll q^2$
(\emph{i.e.} $|t_B|\ll 1$).

Since the gauge theory under consideration has a known string theory
dual description, we will describe the deep inelastic scattering of
an electron from a scalar meson with one flavor in terms of its
holographic dual D3D7-brane model. The ten-dimensional background
metric is
\be
ds^2 = \frac{r^2}{R^2} \, \eta_{\mu\nu} \, dx^\mu dx^\nu +
\frac{R^2}{r^2} \, dr^2 + R^2 \, d\Omega_5^2 \, ,
\ee
where the radius of the five-sphere and the scale of the AdS$_5$
satifies $R^4=4 \pi g_c N \alpha'^2$. The usual four-dimensional
coordinates are $x^\mu=(x^0, \ldots, x^3)$. The induced metric on
the probe D7-brane is given by
\begin{eqnarray}\label{metgral}
ds^{2}_{D7} = \frac{r^2}{R^2} \, \eta_{\mu\nu} \, dx^\mu dx^\nu +
\frac{R^2}{r^2} \, dr^2 + R^2 \, d\Omega^2_{3} \, ,
\end{eqnarray}
which is the asymptotic form of the metric when $r$
is much larger than the distance between the D7-brane and the $N$ D3-branes.
Scalar and vector mesons correspond to excitations of open strings
ending on the probe D7-brane. The dynamics of the D7-brane
fluctuations is described by the DBI action
\begin{eqnarray}\label{kru4}
S_{D7} &=& - T_{7} \int d^{8}\xi \, \sqrt{-\textmd{det}(\hat
P[g]_{ab} + 2 \pi \alpha'F_{ab})}  \, ,
\end{eqnarray}
where $g_{ab}$ stands for the metric (\ref{metgral}),
$T_{7}$
%
%
is the D7-brane tension. In addition, $\hat P$ denotes the
pullback of the background fields on the D7-brane world-volume.

The ten-dimensional $s$-channel Mandelstam variable satisfies
\be
\tilde s \leq - g^{\mu\nu} (P+q)_\mu (P+q)_\nu = \frac{1}{\sqrt{4 \pi g_c N
\alpha'^2}} \bigg(\frac{1}{x}-1\bigg) \, .
\ee

The hadronic tensor can be extracted from the calculation of the
four-point correlation function of two photons and two scalar
mesons. While in the supergravity calculation there is a sum over
intermediate states \cite{Koile:2011aa,Koile:2013hba}, when we focus
on the string theory calculation there is an implicit sum which is
obtained by considering the imaginary part of the forward four-point
scattering amplitude.

In the dual string theory model photons are represented by
gravitons, $h$, with polarizations $h^{M N} = (1/2)(A^M v^N +
A^N v^M)$ where $A^m$ is a $U(1)$ gauge field propagating along $0,
\dots, 4$ coordinates in the bulk\footnote{We use indices $M, N$ from
$0, \dots, 9$; $m, n$ from $0, \dots , 4$; $a, b$ from
$0, \dots , p$; $\mu, \nu$ from $0, \dots , 3$; $i, j$ from $5,
\dots, p$ and $I, J$ from $p+1, \dots, 9$.}. For the D3D7-brane model $v^i$ is
a constant Killing vector on the three-sphere\footnote{For the
D4D6$\mathrm{\overline{D6}}$-brane model and
D4D8$\mathrm{\overline{D8}}$-brane model $v^i$ will be a constant
Killing vector on $S^2$ and $S^4$, respectively.}. In superstring
theory gravitons correspond to closed strings. On the other hand,
scalar mesons are represented by open strings attached to the flavor
D7-brane. Then, one has to calculate the four-point string theory
scattering amplitude of two open and two closed strings, ${\cal
{A}}_4^{2o2c, scalar}$. As mentiones before, this amplitude can be expressed in terms of
two terms with two factors each. One is a kinematic factor, ${\cal
{K}}_i^{2o2c, scalar}$. The other one is a pre-factor ${\cal
{P}}_i^{2o2c, scalar}$ which has the usual gamma function structure,
and it gives the $\alpha'$-dependence:
\bea \label{A2o2c}
{\cal {A}}_4^{2o2c, scalar} = {\cal {P}}_1^{2o2c, scalar} \, {\cal
{K}}_1^{2o2c, scalar} + {\cal {P}}_2^{2o2c, scalar} \, {\cal
{K}}_2^{2o2c, scalar} \, .
\eea
Within the kinematic regime we are interested in the second term
is not relevant. We focus on the first term because
this is the only one having a pole in the $t$-channel \cite{Stieberger:2009hq}.

Since we are interested in the hadronic tensor, we want to calculate
the imaginary part of the forward scattering amplitude. Essentially, we
must evaluate ${\cal {K}}_1^{2o2c, scalar}$ at $\tilde t =0$. This
can be done in two ways. On the one hand, we can take ${\cal
{A}}_4^{2o2c, scalar}$ and replace $h^{m i}$ (the graviton, {\it
i.e.} a closed string) by $A^m v^i$, and also consider $X$
which is the scalar fluctuation represented by an open string. Thus,
in principle, one obtains an effective four-point interaction Lagrangian ${\cal
{L}}_{hhXX}^{eff}$ from which the hadronic tensor can be calculated.
On the other hand, ${\cal {L}}_{hhXX}^{eff}$ is
also given by the coefficient of the $t$-channel graviton pole
multiplied by a pre-factor with the $\alpha'$-dependence. This
second approach can be thought of as the low-energy calculation of
the four-point scattering amplitude, {\it i.e.} in the $\alpha'
\rightarrow 0$ limit. For this calculation one considers the
graviton propagator and the graviton three-point vertex derived from
type IIB supergravity action, and also the graviton-$XX$ vertex
derived from the Dirac-Born-Infeld action with scalar fluctuations
on the flavor brane.

Notice that in what follows the momenta $k_a$ of the fields, the
graviton polarizations $h$, and the polarization of the
vector mesons $\epsilon$ (in Section 3) are parallel to the flavor D7-brane
directions. On the other hand, the scalar mesons represent
fluctuations perpendicular to the flavor D7-brane world-volume
coordinates.

\subsection{Four-point open-closed string theory amplitude}

In order to obtain the tree-level scattering amplitude of two closed
strings representing two gravitons, and two open strings which
represent two scalar (or vector) mesons, we have to carry out a path
integral with the corresponding vertex operator insertions on the
world-sheet. In this case the world-sheet is a disk. In
\cite{Stieberger:2009hq} it has been shown how any amplitude of open
and closed strings can be mapped to pure open string amplitudes\footnote{Note
that these amplitudes were obtained in flat spacetime. We will explain
their relation to curved spacetime amplitudes later.}.
This implies that disk amplitudes, which involve fields in the gauge
multiplets and fields in the supergravity multiplet, are related to
pure amplitudes involving only members from the gauge multiplets.
The world-sheet tree-level diagram of an $S$-matrix for the
open-closed string theory interaction  can be conformally mapped to
a surface with one boundary. Then, following the Riemann mapping
theorem this surface is equivalent to the unit disk $\textbf{D}=\{ z
\in \textbf{C}| , |z| \leq 1 \}$. By using the M\"obius
transformation $z \rightarrow i (1+z)(1-z)^{-1}$ this disk can be
conformally mapped on the upper (complex) half-plane $\textbf{H}_+ =
\{ z \in \textbf{C}| , \textmd{Im}(z) \geq 0 \}$. Then, vertex
operators create the string states corresponding to asymptotic
states in the string theory $S$-matrix formulation. Since in
theories with Dp-branes massless fields correspond to open string
excitations on the Dp-brane world-volume, the disk tree-level
diagram is attached to the Dp-brane world-volume. For closed string
excitations, on the other hand, they propagate in the bulk of the
ten-dimensional spacetime and they are inserted in the interior of
the disk $\textbf{D}$. Thus, the open string vertex operators
$V_o(x_i)$ are inserted at the positions $x_i$ on the boundary of
$\textbf{D}$, where $x_i$ is a real parameter.  On the other hand,
the closed string vertex operators $V_c({\bar {z}_i}, z_i)$ are
inserted at the positions $z_i$ inside the disk $\textbf{D}$.
The open string theory
vertex operators are \footnote{When considering more than one Dp-brane
there is an additional factor which accounts for the non-Abelian
structure. In the present case $N_f=1$ therefore this factor
is just 1.}
\bea
V_{o}^{(-1)} (x, \e_{\mu}, p) &=& g_{o} e^{- \phi} \e_{\mu}
\psi^{\mu} e^{i p\cdot X} (x) \, , \\
V_o^{(0)} (x, \e_{\mu}, p) &=& g_{o} (2 \alpha')^{-1/2} \e_{\mu}
\left(i \partial_x X^{\mu} + 2 \alpha' p\cdot \psi
\psi^{\mu}\right)e^{ip\cdot X}(x) \, ,
\eea
while for closed strings we have
\bea
V_{c}^{(-1,-1)} (z,\bar{z}; h_{\mu \nu}, q) &=& g_{c} e^{-
\widetilde{\phi}(\bar{z})} e^{- \phi(z)}
h_{\mu \nu} \widetilde{\psi}^{\mu}(\bar{z})\psi^{\nu}(z) e^{i q\cdot X (\bar{z},z)} \, , \\
V_c^{(0,0)} (z,\bar{z}; h_{\mu \nu}, p) &=& -g_{c} \frac{2}{\alpha'} h_{\mu \nu}
\left(i \bar{\partial} X^{\mu} + \frac{\alpha'}{2} q\cdot \widetilde{\psi}
\widetilde{\psi}^{\mu}(\bar{z})\right) \left(i \partial X^{\nu}
+ \frac{\alpha'}{2} q\cdot \psi \psi^{\nu}(z)\right) \nn \\
&& e^{iq\cdot X(\bar{z},z) } \, .
\eea
In the present notation $X^{\mu}$ and $\psi^{\mu}$ are the bosonic
and fermionic fields on the world-sheet, while $\phi$ and
$\widetilde{\phi}$ are the ghost fields which come from the
Fadeev-Popov quantization of superstring theory\footnote{Note that
here we use the index notation of \cite{Stieberger:2009hq}.}.
The open and closed string
couplings are $g_o$ and $g_c$, respectively, with $g_c = g_o^2$.

Using these conventions the string theory scattering amplitude
corresponding to two open strings, which are associated with
excitation modes of a flavor Dp-brane, and two closed strings, which
correspond to fields in the bulk, is given by the integral
(\ref{As1}) over the disk. If we were
interested in $1/N$ corrections, the corresponding corrections to
the tree-level string theory scattering amplitudes would be given by
the integrals over different inequivalent topologies. In addition,
in the scattering amplitude we have to include the normal ordering
of each vertex operator. Thus, in what follows whenever we write $V$
we mean $:V:$
\bea \label{As1}
&& A_{string}(h_1, h_4, \e_2, \e_3) = \int_{\bf{\partial H}_+} dx \int_{\bf{\partial H}_+} dy
\int_{z \in \bf{H}_+} dz d\bar{z} \int_{w \in \bf{H}_+} dw d\bar{w} \, \langle c(z)
\widetilde{c}(\bar{z}) \times \nn \\
&& V_c^{(0,0)} (z,\bar{z}; h_{1 \mu \nu}, k_1) V_c^{(-1,-1)}
(w,\bar{w}; h_{4 \mu \nu}, p_4) (c(x) - c (y)) V_o^{(0)} (x, \e_{2\mu}, k_2)
V_o^{(0)} (y, \e_{3\mu}, k_3) \rangle , \nn \\
&&
\eea
where $c$ and $\tilde c$ are ghost fields.
There are two real parameters associated with the insertions of the
open string vertex operators and two complex parameters
corresponding to the insertions of two closed string vertex
operators. The world-sheet $SL(2,R)$ symmetry group allows us to fix
three real parameters. Therefore, the number of integrals reduces to
just three. Thus, a possibility is to set $x \rightarrow - \infty$,
$y = 1$ and $Re(w)=0$ ({\it i.e.} $\bar{w}=-w$). In addition, let us
briefly comment on the vacuum expectation value in Eq.(\ref{As1}),
which corresponds to a path integral over the fields on the
world-sheet which we schematically represent as $\Phi \equiv \{X,
\psi, \widetilde{\psi}, \phi, \widetilde{\phi}, c, \widetilde{c}\}$.
Then, we have
\beq
\langle f[X, \psi, \widetilde{\psi}, \phi, \widetilde{\phi}, c,
\widetilde{c}] \rangle \equiv \int D\Phi \ \ e^{i S[\Phi]} f[\Phi]
\, ,
\eeq
which can be factorized as follows
\beq
S[\Phi] = S[X, \psi, \widetilde{\psi}, \phi, \widetilde{\phi}, c,
\widetilde{c}] = S_{X}[X] + S_{\psi}[\psi,\widetilde{\psi}] +
S_{\phi}[\phi, \widetilde{\phi}] + S_c[c, \widetilde{c}] \, .
\eeq
This implies that each path integral can be done separately. Thus,
in order to calculate the scattering amplitude (\ref{As1}) we can
use the following expectation values of the fields
\cite{Stieberger:2009hq,Hashimoto:1996kf,Fotopoulos:2002wy}
\bea
\langle X^{\mu}(z) X^{\nu}(w) \rangle &=& - 2 \alpha' \eta^{\mu \nu}
\ln (z - w), \nn \\
\langle X^{\mu}(z) \widetilde{X}^{\nu}(\bar{w}) \rangle &=& -2 \alpha'
D^{\mu \nu} \ln (z - \bar{w}), \nn \\
\langle \psi^{\mu}(z) \psi^{\nu} (w) \rangle &=& \eta^{\mu \nu}
(z - w)^{-1}, \nn \\
\langle \psi^{\mu}(z) \widetilde{\psi}^{\nu}(\bar{w}) \rangle &=&
D^{\mu \nu} (z - \bar{w})^{-1}, \nn \\
\langle \phi(z) \phi(w) \rangle &=& - \ln (z - w), \nn \\
\langle \phi(z) \widetilde{\phi}(\bar{w})  \rangle &=& -
\ln (z - \bar{w}), \nn \\
\langle c(w_1) c(w_2) c(z) \rangle &=& C_{ghost} (w_1 - w_2) (w_1 - z)
(w_2 - z), \nn \\
\langle c(w_1) c(w_2) \widetilde{c}(\bar{z})\rangle &=& C_{ghost}
(w_1 - w_2) (w_1 - \bar{z}) (w_2 - \bar{z}), \nn
\eea
where $D^{\mu \nu}$ is a diagonal matrix whose elements are $1$ in
the directions parallel to the flavor Dp-brane and $-1$ in the
perpendicular directions. Thus, we have all the ingredients to
calculate the expectation value in (\ref{As1})
\bea
&&\langle c(z)\widetilde{c}(\bar{z})(c(x)-c(y))\rangle
\langle e^{-\widetilde{\phi}(\bar{z})}e^{-\phi(z)}
\rangle g_c^2 \frac{2}{\alpha'} \, g_o^2\frac{1}{2\alpha'}
h_{1 \mu \nu} h_{4 \rho \sigma} \e_{2 \alpha} \e_{3 \beta} \times  \nonumber \\
&&\langle \widetilde{\psi}^{\mu}(\bar{z})\psi^{\nu}(z)
e^{i k_1\cdot X (\bar{z},z)} \left(i \bar{\partial} X^{\rho}
+ \frac{\alpha'}{2} k_4\cdot \widetilde{\psi}
\widetilde{\psi}^{\rho}(\bar{w})\right) \left(i \partial X^{\sigma}
+ \frac{\alpha'}{2} k_4 \cdot \psi \psi^{\sigma}(w)\right)
e^{ik_4\cdot X(\bar{w},w) } \nonumber \\
&& \left(i \partial_x X^{\alpha} + 2 \alpha' k_2\cdot \psi
\psi^{\alpha}\right)e^{ik_2\cdot X}(x)\left(i \partial_y X^{\beta} +
2 \alpha' k_3\cdot \psi \psi^{\beta}\right)e^{ik_3\cdot X}(y)
\rangle.
\eea
We have to obtain all the Wick's contractions for 16 different
terms. An important simplification comes from the fact that the
contraction of two fields at the same point on the disk vanishes.
The calculation is rather complicated but in the case of scalar
mesons there is only one non-vanishing term.  This is due to the
fact that its corresponding polarizations (which are non-vanishing
only in the perpendicular directions to the Dp-brane) are themselves
perpendicular to all momenta as well as to all the rest of the
polarizations of the fields. An early result was obtained by
Fotopoulos and Tseytlin in \cite{Fotopoulos:2002wy} within the
regime where superstring theory can be described by supergravity. In
this case the integral was carried out close to the singularities.
We perform similar calculations in the following section. More
recently, in an extensive work Stieberger obtained the exact result
\cite{Stieberger:2009hq}, which can be split into two terms as
anticipated in Eq.(\ref{A2o2c}), where
\bea
{\cal {P}}_1^{2o2c, scalar} &=& \int_{-\infty}^{+\infty} dx \,
x^2 (1+ix)^{u-1} (1-ix)^{u-1} \times \label{Prefactor1}\\
&& \int_{\mathbb{C}} d^2z \, (1-z)^{s} (1-\bar{z})^{s}
(z+ix)^{\frac{t}{2}-1}(z-ix)^{\frac{t}{2}-1}
(\bar{z}+ix)^{\frac{t}{2}-1}(\bar{z}-ix)^{\frac{t}{2}-1},
\nn
\eea
while ${\cal {K}}_1^{2o2c, scalar}$  can be obtained as the sum of
the scattering
amplitudes associted with the different Feynman diagrams from the
supergravity calculation \cite{Fotopoulos:2002wy}. We only write the
first term because in the kinematic regime which we are interested
in only this term is relevant.

The pre-factor needed in order to construct the effective action 
of two closed-two open strings interaction is formally given by 
the small $\tilde{t}$ and large $\tilde{s}$ limit of expression 
(\ref{Prefactor1})\footnote{Recall that as we are dealing with 
massless particles $\tilde{s} + \tilde{t} + \tilde{u} = 0$, 
here the absolute value of $\tilde{u}\approx -\tilde{s}$ also 
becomes large.}. However, since it is rather difficult to deal 
with we will give several arguments to support the assumption that, 
in the scalar case, this pre-factor takes the same form that it 
has in the case of glueballs, \textit{i.e.} 
\beq
\frac{\pi \alpha'}{4} \sum_{m=0}^{\infty} \delta 
\left(m - \frac{\alpha' \tilde{s}}{4}\right) (m)^{\alpha' \tilde{t}/4},
\eeq
where we have omitted the $1/\tilde{t}$ term which leads to the 
$t$-channel as the dominant contribution and focused on the imaginary 
part that singles out the exchange of excited strings \cite{Polchinski:2002jw}. 
In \cite{Brower:2006ea} it has been shown that in this parameter regime 
the $\tilde{s}$-dependence always gives a factor 
\beq
\left(\frac{\alpha' \tilde{s}}{4}\right)^{\alpha' \tilde{t}/4} = 
\int_0^{\infty} dm \, \delta \left(m - \frac{\alpha' 
\tilde{s}}{4}\right) (m)^{\alpha' \tilde{t}/4} \approx  
\sum_{m=1}^{\infty} \delta \left(m - \frac{\alpha' \tilde{s}}{4}\right) 
(m)^{\alpha' \tilde{t}/4}.
\eeq
In the last step we have considered the fact that when $x \ll 1/\sqrt{\lambda}$ 
the integral and the sum are not very different. Note that in the small $x$ regime the factor
$(m)^{\alpha' \tilde{t}/4}$ is order one \cite{Polchinski:2002jw}. 
This approximation breaks down in the exponentially small regime.
There is also a factor carrying the pole in $\tilde{t}$. 
In fact, the OPE expansion of the operators involved in this process have been 
studied both for a pair of closed superstrings \cite{Brower:2006ea} and for open superstrings  
\cite{Cheung:2010vn,Fotopoulos:2010cm}. In both cases it is shown that 
this function gives a factor of the form 
\beq
\frac{\Gamma\left(-1-\frac{\alpha' \tilde{t}}{4}\right)}{\Gamma\left(2
+\frac{\alpha' \tilde{t}}{4}\right)} e^{i\pi(1-\alpha't/4)} \propto  
-\frac{1}{\alpha' \tilde{t}} + O(1),
\eeq
where the last result comes from the small $\tilde {t}$ expansion. 
Therefore, this supports the fact that we can neglect other 
possible terms in ${\cal {A}}_4^{2o2c, scalar}$ and ${\cal {A}}_4^{2o2c, vector}$, 
besides the one which does not vanish in the $\alpha' \rightarrow 0$ limit.  
In addition, the OPE analysis also singles out the term where the 
polarizations of incoming gravitons and mesons ($h$ and $\e$) are contracted 
among themselves as follows: $(h_1h_4)(\e_2\e_3)$. As we will see, this is 
in complete agreement with our calculations.

Finally,  the assumption about the pre-factor is also supported
\textit{a posteriori}  by our results: as will be demonstrated 
in the following sections, the kinematic part of the effective Lagrangian 
in the scalar meson case is identical to the one of \cite{Polchinski:2002jw} 
for glueballs. In addition, we have seen that from the string theory point 
of view the initial vertex operator integral, which leads to the full scattering 
amplitude at genus zero, is the same for scalar and vector mesons. This is because the only 
difference is given by the polarization vectors. This fact suggests the use of the same 
pre-factor as for the case of glueballs.

\subsection{Four-point graviton-scalar meson tree-level amplitudes
from supergravity}

\begin{figure}
\unitlength=1mm
\begin{center}
\begin{fmffile}{Schannel2} 
\begin{fmfgraph*}(40,25)
\fmfleft{i1,i2}
\fmfright{o1,o2}
\fmf{dashes}{i1,v1}
\fmf{photon}{i2,v1}
\fmf{dashes}{v2,o1}
\fmf{photon}{v2,o2}
\fmf{dashes}{v1,v2}
\fmfdot{v1,v2}
\fmflabel{$h_1$}{i2}
\fmflabel{$h_4$}{o2}
\fmflabel{$X_2$}{i1}
\fmflabel{$X_3$}{o1}
			 \marrowup{a}{up}{top}{$k_1$}{i2,v1}
       \marrowup{b}{down}{bot}{$k_2$}{i1,v1}
       \marrowup{d}{up}{top}{$k_4$}{o2,v2}
       \marrowup{e}{down}{bot}{$k_3$}{o1,v2}
\end{fmfgraph*}
\end{fmffile}
\hspace{15pt}
\begin{fmffile}{Uchannel2}
\begin{fmfgraph*}(40,25)
\fmfleft{i1,i2}
\fmfright{o1,o2}
\fmf{phantom}{i2,v1}
\fmf{phantom}{v2,o2}
\fmf{dashes}{v1,i1}
\fmf{dashes}{v2,o1}
\fmf{dashes}{v1,v2}
\fmf{photon,tension=0}{v1,o2}
\fmf{photon,tension=0}{v2,i2}
\fmfdot{v1,v2}
\fmflabel{$h_1$}{i2}
\fmflabel{$h_4$}{o2}
\fmflabel{$X_2$}{i1}
\fmflabel{$X_3$}{o1}
			 \marrowshort{a}{up}{top}{$k_1$}{i2,v2}
       \marrowup{b}{down}{bot}{$k_2$}{i1,v1}
       \marrowshort{d}{up}{top}{$k_4$}{o2,v1}
       \marrowup{e}{down}{bot}{$k_3$}{o1,v2}
\end{fmfgraph*}
\end{fmffile}

\vspace{35pt}

\begin{fmffile}{Tchannel2}
\begin{fmfgraph*}(40,25)
\fmfleft{i1,i2}
\fmfright{o1,o2}
\fmf{photon}{i2,v1,o2}
\fmf{dashes}{i1,v2,o1}
\fmf{photon}{v1,v2}
\fmfdot{v1,v2}
\fmflabel{$h_1$}{i2}
\fmflabel{$h_4$}{o2}
\fmflabel{$X_2$}{i1}
\fmflabel{$X_3$}{o1}
      \marrow{a}{up}{top}{$k_1$}{i2,v1}
       \marrow{b}{up}{top}{$k_4$}{o2,v1}
       \marrow{d}{down}{bot}{$k_3$}{o1,v2}
       \marrow{e}{down}{bot}{$k_2$}{i1,v2}
\end{fmfgraph*}
\end{fmffile}
\hspace{15pt}
\begin{fmffile}{Contact2}
\begin{fmfgraph*}(40,25)
\fmfleft{i1,i2}
\fmfright{o1,o2}
\fmf{photon}{i2,v1,o2}
\fmf{dashes}{i1,v1,o1}
\fmfdot{v1}
\fmflabel{$h_1$}{i2}
\fmflabel{$h_4$}{o2}
\fmflabel{$X_2$}{i1}
\fmflabel{$X_3$}{o1}
			 \marrowup{a}{up}{top}{$k_1$}{i2,v1}
       \marrowup{b}{up}{top}{$k_4$}{o2,v1}
       \marrowup{d}{down}{bot}{$k_3$}{o1,v1}
       \marrowup{e}{down}{bot}{$k_2$}{i1,v1}
\end{fmfgraph*}
\end{fmffile}
\vspace{10pt}
{\caption{\small The four Feynman diagrams corresponding to the
holographic calculation of the four-point amplitude: $s$-, $u$-, 
and $t$-channels plus the contact interaction. Wavy lines
represent gravitons, $h_1$, $h_4$ and the off-shell one. Dashed lines represent scalar
mesons, $X_2$ and $X_3$. $k_i$ indicates the momenta of the fields. }} 
\label{FigScalarSugra}
\end{center}
\end{figure}
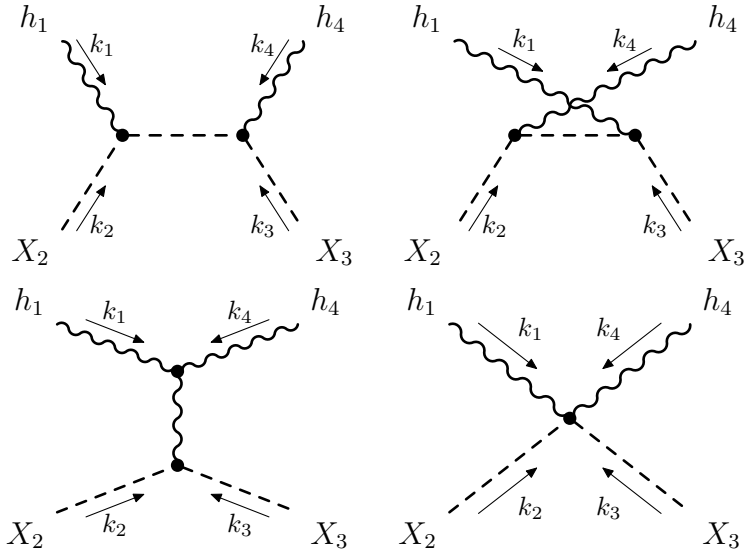
\unitlength=1pt

We begin with the action
\beq
S_{DBI}=-T_7 \int d^8 \xi \sqrt{- det\left(\hat{P}[g]_{ab} + (2\pi
\alpha')F_{ab}\right)} \, ,
\eeq
and
\beq
\sqrt{- g} \approx \sqrt{-g_0} \left[ 1 + \frac{1}{2} H -
\frac{1}{4}H^a_b H^b_a + \frac{1}{8} H^2 + \frac{1}{6}H^a_b H^b_c
H^c_a - \frac{1}{8} H H^a_b H^b_a + \frac{1}{48} H^3\right] \, ,
\label{taylor}
\eeq
where $H \equiv H^a_a$ stands for trace of this tensor field. There are eight
coordinates which are parallel to the D7-brane, and two perpendicular ones, 
$x^8$ and $x^9$. We can describe the degrees of freedom of the system by
identifying $\xi^{a} \equiv X^{a}$ with $a = 0, \dots, 7$ and
reinterpreting the $(8, 9)$-coordinates as scalar fields $X^I$ with
$I =1, 2$. Their variation parametrize fluctuations of the D7-brane
along its normal directions. By using this static parametrization,
and by ignoring all vector fields $F_{ab}$, we identify the fields
in our theory with the metric perturbations associated with the
graviton and the scalar fields as follows
\ba
H_{ab} \rightarrow \hat{g}_{ab} \equiv g_{ab} + 2 g_{I(a}\partial_{b)}X^{I}
+ g_{IJ} \, \partial_aX^{I}\partial_{b}X^{J}, \\
g_{ab} = \eta_{ab} + 2 \kappa h_{ab} \ , \ g_{IJ} = \delta_{IJ}
+ 2\kappa h_{IJ} \ , \ g_{aI} = 2\kappa h_{aI}.
\ea
Now, we will focus on the case where the graviton polarization is
parallel to the D7-brane, which implies that $h_{IJ} = h_{aI} = 0$.
By expanding $S_{DBI}$,  we obtain three of the necessary
ingredients for the calculation of the tree-level Feynman diagrams.
These diagrams are analogous to those obtained in the low-energy
limit of a closed string theory, when considering graviton-dilaton
interactions, namely: the kinetic term associated with the scalar
fields and the interactions of the type $hXX$ and $hhXX$. Notice
that since the scalar term in $H$ is quadratic, we only need to do
the expansion up to third order. Also recall that the three-graviton
vertex and the graviton propagator come directly from the closed
string theory action
\cite{Gross:1986mw,Sannan:1986tz,Fotopoulos:2002wy} \footnote{This
is so because the corrections coming from $S_{DBI}$ are
sub-leading.}.

The first order in the expansion above gives a kinetic term
$\frac{T_7}{2} \partial_a X \partial^a X$ \footnote{From now on, we
drop indices $I$ and $J$ whenever they are summed, by using the
Kronecker delta $\delta_{IJ}$.}. Therefore, the propagator is
\beq
P_{IJ}(k) \equiv \langle X_I X_{J} \rangle = - i \frac{\delta_{IJ}}{T_7 k^2}.
\eeq
The second and third orders in the expansion produce the following
interaction Lagrangians
\ba
L_{hXX} & = & T_7 \kappa \left[\frac{1}{2} h \partial^a X \partial_a X
- h^a_b \partial^bX \partial_a X\right] ,\\
L_{hhXX} & = & T_7 \kappa^2 \left[ \frac{1}{4} h^2 \partial^a X \partial_a X
+ 2 h^a_b h^b_c \partial^c X \partial_a X  - h h^a_b \partial^b X
\partial_a X - \frac{1}{2} h^a_b h^b_a \partial^c X \partial_c X\right], \nn \\
&&
\ea
where $h = h_{a}^a$. Notice that for the
contact term $V_{hhXX}$ we can ignore terms with a factor $h^a_a$
since the external gravitons are on-shell and therefore traceless.
However, this does not hold for the $V_{hXX}$ vertex since in the
$t$-channel diagram this vertex exchanges a virtual graviton with
the three-graviton vertex\footnote{In order to write down the
three-graviton vertex coupled to an off-shell graviton one has to
consider that in the expansion above $h_{IJ}$ does not vanish. This
adds an extra term which however does not contribute due to the
graviton propagator structure \cite{Fotopoulos:2002wy}.}. Let us
call $k_1$ and $k_2$ the incoming momenta of the scalar field, then we obtain
\ba
V_{hXX}^{ab,IJ} &=& i T_7 \kappa \, \delta^{IJ}\left[\frac{1}{2}
\eta^{ab} (k_1\cdot k_2)
-(k_1^a k_2^b + k_1^bk_2^a) + \text{irrelevant}\right] \, \\
V_{hhXX}^{ab, cd, IJ}(k_1, k_2) &=& 4 i T_7 \kappa^2 \, \delta^{IJ}
\left[ \eta^{bc} (k_1^a k_2^d + k_1^dk_2^a)
- \frac{1}{2} \eta^{ac}\eta^{bd} (k_1 \cdot k_2)\right] \, .
\ea
For instance, the scattering amplitude related to the contact
diagram (subindex $c$ indicates contact term) is given by
\ba
A_c(k_1, k_2, k_3, k_4) &=& h_{1ab}\, h_{4de} \, \delta_{IJ}\,
V_{hhXX}^{ab,de,IJ}(k_2,k_3)  \nn \\
&=& i \kappa^2 T_7  \left[ 4 (k_2 h_1 h_4 k_3)
+ 4 (k_3 h_1 h_4 k_2) - 2 (h_1 h_4) (k_2 \cdot k_3)\right], 
\ea
where indices of
factors within parentheses are totally contracted. In addition, in
order to calculate the scattering amplitudes corresponding to the
$s$- and $u$-channels from the previous vertices, and by using the
transversal character of the polarizations, momentum conservation
and the dispersion relations for massless particles, we obtain the
following results\footnote{Notice that whenever we write $s$, $t$,
$u$ in the string theory scattering amplitudes we actually mean $\tilde s$, $\tilde t$,
$\tilde u$, since these are the actual ten-dimensional Mandelstam variables.}
\ba
A_s(k_1,k_2,k_3,k_4) &=& h_{1ab}\, h_{4cd} \, \delta_{IJ} \,
V_{hXX}^{ab,IK}(k_2,-k_1-k_2) P_{KL}(-k_1-k_2) V^{cd,LJ}(k_1 + k_2,k_3) \nn \\
&=&  - i \kappa^2 T_7 \frac{4}{s} (k_2 h_1 k_2)(k_3 h_4 k_3) \, , \\
A_u(k_1,k_2,k_3,k_4) &=& h_{1ab}\, h_{4cd} \, \delta_{IJ} \,
V_{hXX}^{ab,JK}(k_3,-k_1-k_3) P_{KL}(-k_1-k_3) V^{cd,LI}(k_1 + k_3,k_2) \nn \\
&=& - i \kappa^2 T_7 \frac{4}{u} (k_3 h_1 k_3)(k_2 h_4 k_2) \, ,
\ea
where $s=-2k_1\cdot k_2 = -2 k_3 \cdot k_4$ while $u =-2k_1\cdot k_3 =
-2 k_2 \cdot k_4$.

Now, let us consider the $t$-channel. We need three pieces: the
three-graviton vertex derived from $L_{hhh}$, the graviton
propagator which is derived from $L^{\partial^2}_{hh}$, and the
three-point interaction vertex with one graviton and two scalars
derived from $L_{hXX}$ which we have already obtained.
The graviton that connects the three-graviton vertex from
$L_{hhh}$ and the interaction vertex $V_{hXX}$ is off-shell, and
therefore we cannot neglect the first term. Also, note that since we
will contract this vertex with the graviton propagator it is not
necessary to symmetrize the vertex in the $\alpha$ and $\beta$ Lorentz
indices.  This vertex must then be contracted with a factor
$S^{\rho \sigma, \gamma \delta, \alpha \beta}(k_1,k_4)$ obtained for
example by Sannan \cite{Sannan:1986tz}, which corresponds to the
contraction of the off-shell graviton propagator and the
three-graviton vertex. We must also contract this factor with the
polarizations of the external gravitons $h_1$ and $h_4$. The
resulting factor is
\bea
&& h_{1 \rho \sigma} h_{4 \gamma \delta} S^{\rho \sigma, \gamma
\delta, \alpha \beta}(k_1,k_4) = \nonumber \\
&& - \kappa \left[ \frac{1}{2}(h_1h_4)^{\alpha \beta} +
\frac{1}{2}(h_1h_4)^{\beta \alpha} +
\frac{1}{t}\left((k_1h_4)^{\alpha}(k_4 h_1)^{\beta} +
(k_1h_4)^{\beta}(k_4 h_1)^{\alpha} \right) \nonumber \right. \\
&& \left. - \frac{(h_1h_4)}{t} \left( k_1^{\alpha} k_1^{\beta} +
k_4^{\alpha} k_4^{\beta} + \frac{1}{2}k_4^{\alpha} k_1^{\beta} +
\frac{1}{2}k_1^{\alpha} k_4^{\beta}\right)
- \frac{1}{t}\left((k_1 h_4 k_1)h_1^{\alpha \beta}
+ (k_4 h_1 k_4) h_4^{\alpha \beta}\right) \nonumber \right. \\
&& \left. + \frac{1}{t}    \left((k_1 h_4 h_1)^{\alpha}k_1^{\beta} +
(k_1 h_4 h_1)^{\beta}k_1^{\alpha} + (k_4 h_1
h_4)^{\alpha}k_4^{\beta} + (k_4 h_1 h_4)^{\beta}k_4^{\alpha} \right)
\right]. \label{eq-sanann}
\eea
As mentioned before we take $\alpha,\beta \rightarrow a,b $ and
contract this result with the vertex $V_{hXX}^{ab , I, J}$,
obtaining the $t$-channel amplitude for the process
\ba
&&A_t(k_1,k_2,k_3,k_4) = h_{1 cd} h_{4 ef} S^{cd, df, ab}(k_1,k_4)
V_{hXX}^{ab, I, J}
(k_2,k_3) \delta_{IJ} = \nn \\
&&   \frac{- 2i\kappa^2 T_7}{sut} \big[ \frac{1}{4}
(h_1 h_4) (t^2 + u^2 + s^2)  + u [(k_3h_1h_4k_1) + (k_4h_1h_4k_2)]  \\
&&+ s [(k_2h_1h_4k_1) + (k_4h_1h_4k_3)] + t [(k_2h_1h_4k_3) +
(k_3h_1h_4k_2) - (k_4h_1h_4k_1)] + \nn \\
&&  2\left[ (k_4 h_1 k_2)(k_3 h_4 k_1) + (k_4 h_1 k_3)(k_2 h_4 k_1) -
(k_4 h_1 k_4)(k_2 h_4 k_3) - (k_2 h_1 k_3)(k_1 h_4 k_1) \right] \big], \nn
\ea
being $t = -2 k_1 \cdot k_4 = -2 k_2 \cdot k_3$. Notice that
$(su)^{-1}$ has been factorized for convenience, however the
graviton pole of the $t$-channel does appear.

The total tree-level scattering amplitude associated with this
process is $A_{total} = A_c + A_s + A_u + A_t$. As we shall see
later, this structure coincides with the scattering amplitude of
closed string theory with two gravitons and two dilatons up to a
global factor $su$. This means that, since the Feynman diagrams
are the same, $A_{total}$ reproduces the
kinematic factor of the closed string theory four-point scattering
amplitude obtained from the world-sheet integration of the
expectation value of four string theory vertex operators described
in the previous section. In fact, the four-point closed string
scattering amplitude is a known result given by
\beq
A_4^{closed} = - i \pi^2 g_{c} \alpha'^3
\frac{\Gamma\left(-\frac{\alpha's}{4}\right)\Gamma\left(-\frac{\alpha'u}{4}\right)
\Gamma\left(-\frac{\alpha't}{4}\right)}{\Gamma\left(1+\frac{\alpha's}{4}\right)
\Gamma\left(1+\frac{\alpha'u}{4}\right)
\Gamma\left(1+\frac{\alpha't}{4}\right)} \times K_4^{closed} \, ,
\eeq
where the pre-factor is proportional to $(stu)^{-1}$ at first order
in $\alpha'$. On the other hand, $K_4^{closed}$ is a kinematic
factor which contains the polarizations. This does not depend on
$\alpha'$ and can be separated into kinematic factors associated
with open string theory scattering amplitudes as
\beq
K_{4}^{closed}(k_1, k_2, k_3, k_4) \propto K_{4}^{open}
\left(\frac{1}{2}k_1, \frac{1}{2}k_2, \frac{1}{2}k_3,
\frac{1}{2}k_4 \right) \times K_{4}^{open}\left(\frac{1}{2}k_1,
\frac{1}{2}k_2, \frac{1}{2}k_3, \frac{1}{2}k_4 \right).
\eeq
We can explicitly calculate the scattering amplitude\footnote{We
have done this calculation of the string theory scattering
amplitudes by using a field-theory motivated approach to symbolic
computer algebra called Cadabra
\cite{Peeters:2006kp,Peeters:2007wn}.} from $K_4^{open}$ which can
be found in \cite{Green:1987sp}, and replace the graviton
polarizations as $h_1^{MN}$ and $h_2^{MN}$, using the
transversality condition and the fact that on-shell gravitons are
traceless. The dilatons' polarizations are also transversal
\ba
\frac{1}{\sqrt{8}} \phi_2 \left(\eta^{MN} - k_2^{M}\bar{k_2}^{N}
- k_2^{N}\bar{k_2}^{M}\right) \ ,
\ \frac{1}{\sqrt{8}} \phi_3 \left(\eta^{MN} - k_3^{M}\bar{k_3}^{N}
- k_3^{N}\bar{k_3}^{M}\right). \nonumber
\ea
The auxiliary momenta $\bar{k_2}$ and $\bar{k_3}$ satisfy the
following relations $\bar{k_2} \cdot \bar{k_2} = \bar{k_3}\cdot
\bar{k_3} = 0$ and $k_2 \cdot \bar{k_2} = k_3 \cdot \bar{k_3} = 1$.
The only reason to include them is in order to have transversal
polarizations. Thus, although they are important within the
intermediate steps in the explicit calculations they never appear in
the final results. Therefore, it is said that these momenta decouple
from any physical process.

Recall that we want to study string theory scattering amplitudes in
a curved spacetime. It is then important to think of the validity of
the calculations that we perform since we do it for the string
theory scattering amplitudes in flat spacetime. In order to answer
this question we may separate the fields of the conformal theory on
the world-sheet from the interaction, by considering the zero modes
separated from the excited modes, {\it  i.e.} $X^{M}(\tau, \sigma) =
x^{M} + \tilde{X}^{M}(\tau, \sigma)$. At a
fixed point $x^{M}$, the Gaussian integral over the modes
$\tilde{X}^{M}$ is exactly the same as the one we would have
in flat spacetime\footnote{The Ramond-Ramond fields induce
perturbations which are sub-leading in the present regime, with
$\frac{R^2}{\alpha'} = \sqrt{\lambda} >> 1$, therefore we do not
need to consider them.} rendering a local amplitude $i
A_{local}(x^{M})$. Thus, if we integrate over the zero modes, {\it i.e.}
in flat spacetime, we obtain an $S$-matrix of the form
\beq
S = i \int d^{10}x \, \sqrt{-g} \, A_{local}(x^{M}).
\eeq
This is equivalent to say that the interaction is approximately local, and
this holds because the scale $\alpha'$ is small compared to the
spacetime curvature. This {\it local} approximation breaks down when
$x$ becomes exponentially small \cite{Brower:2006ea}.

\subsection{Effective Lagrangian with four-point interactions}

The idea now is to obtain an effective action with four-point
interactions involving two gravitons and two scalar mesons. Thus, we
have to start from the four-point string theory scattering amplitude
that we have discussed before. It is important to recall that since
the Bjorken parameter is small in the DIS regime we consider, {\it
i.e.} $t \approx 0$ is very small, the Mandelstam variables $u$ and
$s$ become very large. On the one hand, it is necessary to perform
appropriate approximations on the pre-factor which we will discuss
later. On the other hand, we have to find the leading terms coming
from the kinematic factor, which are those that produce the
$t$-channel scattering amplitude in the supergravity approximation.
Since we look for a four-point interaction Lagrangian we can start
by writing a list of the terms which, {\it a posteriori}, will be
necessary in order to reproduce the $t$-channel scattering amplitude
at leading order in the $t\approx 0$ approximation. These are
\ba
{\cal {L}}_{A} &=&\partial_n h_{mp} \partial^n h^{mp}\partial_M X
\partial^M X  \rightarrow  (h_1h_4) t^2, \label{traza1} \nn \\
{\cal {L}}_{B} &=&\partial^m h^{pr} \partial^n h_{pr} \partial_m X
\partial_n X   \rightarrow\frac{1}{2}   (h_1h_4) (s^2 + u^2),
\label{traza2}\nn \\
{\cal {L}}_{C} &=&\partial_m h^{np} \partial_n h_p^m \partial_MX
\partial^M X \rightarrow -2 t   (k_4 k_1 h_4 k_1),
\label{khhk1} \nn \\
{\cal {L}}_{D} &=& \partial^m h^{pr} \partial_p h_r^n\partial_m X
\partial_n X \rightarrow -\frac{1}{2}s  \left[(k_4 h_1 h_4k_3)
+ (k_2 h_1 h_4 k_1)\right] \label{khhk2} \nonumber \nn \\
&& -\frac{1}{2}u  \left[ (k_3 h_1 h_4 k_1)
+ (k_4 h_1 h_4 k_2)\right],  \nn \\
{\cal {L}}_{E} &=&\partial^p h^{mr} \partial_p h_r^n \partial_m X
\partial_n X \rightarrow  - t  [(k_2 h_1 h_4 k_3)
+ (k_3 h_1 h_4 k_2) ], \label{khhk3}\nn \\
{\cal {L}}_{F} &=&\partial_p h^{rm} \partial_r h^{pn} \partial_m X
\partial_n X   \rightarrow 2  [ (k_2 h_1 k_4)(k_3 h_4 k_1)
+ (k_3 h_1 k_4)(k_2 h_4 k_1) ], \label{khk1}\nn \\
{\cal {L}}_{G} &=&h^{rp}\partial_r \partial_p h^{mn} \partial_m X
\partial_n X  \rightarrow 2  [(k_1 h_4 k_1)(k_2 h_1 k_3)
+ (k_4 h_1 k_4)(k_2 h_4 k_3)]. \label{khk2}
\ea
By comparison with the coefficient of the $t$-channel graviton pole
we can write the following effective Lagrangian
\beq
{\cal {L}}_{eff} \propto \frac{1}{4} \times {\cal {L}}_{A} +
\frac{1}{2} \times {\cal {L}}_{B} + \frac{1}{2}\times {\cal {L}}_{C}
- 2 \times {\cal {L}}_{D} - 1 \times {\cal {L}}_{E} + 1\times {\cal
{L}}_{F} - 1 \times {\cal {L}}_{G}.
\eeq
There is also a global factor $s u$ which, for convenience, we will
include in the pre-factor.
Now, the metric perturbation we consider is $h^{ab} \equiv
\frac{1}{2}(A^a v^{b} + A^b v^a)$, where $A^m$ is a $U(1)$
gauge field  while $v^i$ is a constant
Killing vector on $S^3$. Recall that the D7-brane wraps this sphere,
therefore we can rewrite the effective Lagrangian using the
following identities
\ba
\partial_b h_{ac} \partial^b h^{ac} &=& \frac{1}{2} v^cv_c
\partial_b A_a \partial^b A^a \ , \ \partial^a h^{cd}
\partial^b h_{cd} = \frac{1}{2} v^cv_c \partial^a A_d
\partial^b A^d, \nn \\
\partial_c h^{da} \partial_d h^{cb} &=& \frac{1}{4} v^av^b
\partial_c A^d \partial_d A^c \ , \ \partial_a h^{bc}
\partial_b h_c^a = \frac{1}{4} v^cv_c \partial_a A^b
\partial_b A^a,  \nn \\
\partial^a h^{cd} \partial_c h_d^b &=& \frac{1}{4} v^d v_d
\partial^a A^c \partial_c A^b \ , \ \partial^c h^{ad}
\partial_c h_d^b  = \frac{1}{4} ( v^dv_d \partial^c A^a
\partial_c A^b + v^a v^b \partial^c A^d \partial_c A_d) . \nn
\ea
Once a graviton polarization is chosen in this way the term
${\cal{L}}_{G}$ vanishes. By taking into account that the terms in
the scattering amplitude which are proportional to $t$ are
irrelevant in the present regime, we can write an explicitly gauge
invariant effective Lagrangian for the gauge field $A^m$. Thus, we
have
\ba
{\cal {L}}_{eff} \propto \frac{1}{4} \left[2 \, v^i v_i \,
\partial_{m}X \partial^p X \, F^{mn} F_{pn} + v^i v^j \, \partial_i X \partial_j X
\, F^{mn} F_{mn} \right] \, , \label{effLagscalar}
\ea
where $F_{mn} = \partial_m A_n - \partial_n A_m$. In this
Lagrangian\footnote{We have
explicitly checked that there is a misprint in the
sign of the second term in Eq.(82) in reference  \cite{Polchinski:2002jw}. We have
done the explicit calculations both for glueballs as in
\cite{Polchinski:2002jw} and for scalar mesons in two different ways
and we have confirmed our results.  However,
fortunately these terms do not contribute to the calculation of the
structure functions. Neither we include a third term which goes like
$t$ which is present in Eq.(82) of reference
\cite{Polchinski:2002jw}.} the most important term is the first one
since it leads to terms proportional to $s^2$ and $s$ in the
scattering amplitude. As we shall see in Section
3, there is an analogous effective Lagrangian for vector mesons.

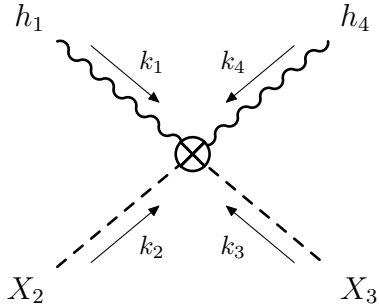
\begin{figure}
\unitlength=1mm
\begin{center}
\begin{fmffile}{Eff2}
\begin{fmfgraph*}(45,30)
\fmfcmd{
    path quadrant, q[], otimes;
    quadrant = (0, 0) -- (0.5, 0) & quartercircle & (0, 0.5) -- (0, 0);
    for i=1 upto 4: q[i] = quadrant rotated (45 + 90*i); endfor
    otimes = q[1] & q[2] & q[3] & q[4] -- cycle;
}
\fmfwizard
\fmfleft{i1,i2}
\fmfright{o1,o2}
\fmf{photon}{i2,v1,o2}
\fmf{dashes}{i1,v1,o1}
\fmfv{d.sh=otimes,d.f=empty,d.si=.1w}{v1}
\fmflabel{$h_1$}{i2}
\fmflabel{$h_4$}{o2}
\fmflabel{$X_2$}{i1}
\fmflabel{$X_3$}{o1}
       \marrowup{a}{up}{top}{$k_1$}{i2,v1}
       \marrowup{b}{up}{top}{$k_4$}{o2,v1}
       \marrowup{d}{down}{bot}{$k_3$}{o1,v1}
       \marrowup{e}{down}{bot}{$k_2$}{i1,v1}
\end{fmfgraph*}
\end{fmffile}
\vspace{10pt}
{\caption{\small Effective four-point interaction. Wavy lines
represent on-shell gravitons. Dashed lines represent scalar
mesons.}} \label{FigEffScalar}
\end{center}
\end{figure}
\unitlength=1pt

\subsection{Hadronic tensor at small $x$ for scalar mesons}

We can use the same approximation as in \cite{Polchinski:2002jw},
namely: based on the the effective Lagrangian (\ref{effLagscalar})
we can construct an effective action in curved spacetime by
contracting all indices with the ten-dimensional metric in curved
spacetime and also by multiplying by the squared root of the
determinant of that metric as usual. Therefore, we begin
with\footnote{Notice that within the present kinematic regime,
where $1/\exp{\sqrt{\lambda}} \ll x \ll 1/\sqrt{\lambda}$, $\tilde {s}$ is
taken as a scalar quantity and not as a differential operator
\cite{Polchinski:2002jw}.}
\beq
n^*_{\mu} n_{\nu} \text{Im}_{exc}T^{\mu \nu} = \frac{\pi \alpha'}{8}
\sum_{m=1}^{\infty} \int d\Omega_3 \, dr \, \sqrt{-g} v_{i}v^{i}
\partial^{k} X^{*}(P)\partial^nX(P)F_{kp}^{*}(q)F_{n}^{p}(q)
\delta\left(m-\frac{\alpha' \tilde {s}}{4}\right) \, . \label{iTmunu}
\eeq
We have omitted the integrals in the four-dimensional spacetime
$x_0, \cdot \cdot \cdot, x_3$ since we have already used momentum
conservation. The solutions to the field equations are \cite{Koile:2011aa}
\bea
A_{\mu}(q) &=& n_{\mu} f(r) e^{i q\cdot x} \, , \\
A_{r} (q) &=& \frac{-i q\cdot n}{q^2} f'(r) e^{i q\cdot x}  \, , \\
X^l &=& \frac{c_i^l}{\Lambda R^3} \left(\frac{r}{\Lambda R^2}\right)^{-\Delta}
Y^l(\Omega_3) e^{i P\cdot x} \, ,
\eea
where $\Delta$ stands for the scaling dimension of the meson fields and 
$c_i$ is a dimensionless constant.
By using $\w = q R^2/r$ we can write $f = \w K_1(\w)$, where
$K_1$ is the modified Bessel function of the second kind. In order
to simplify the notation we drop the label $l$ which is associated
with the spherical harmonic. Recall that this index identifies each
meson. The corresponding field strengths are
\bea
F_{\mu \nu} &=& i (q_{\mu}A_{\nu} - q_{\nu} A_{\mu}) = i (q_{\mu}n_{\nu}
- q_{\nu} n_{\mu}) f(r) e^{i q\cdot x} \, , \\
F_{\mu r} &=& i\left(q_{\mu} - n_{\mu} \frac{q^2}{q\cdot n}\right) A_r
= \left( \frac{q \cdot n}{q^2} q_{\mu} - n_{\mu}\right) f'(r)
e^{i q \cdot x} \, , \\
\partial_{\nu}X &=& i P_{\nu} X \, , \,\,\,\,\,\,  \partial_{r}X
= \frac{-\Delta}{r} X \, ,
\eea
from which we can construct the different pieces involved in the
effective Lagrangian
\bea
F_{rp}^{*}F_r^p &=& \frac{R^2}{r^2} \eta^{\rho \sigma} F_{r\rho}^{*}
F_{r \sigma}
= \frac{R^2}{r^2} (f')^2 \left[(n \cdot n^{*})
- \frac{(q\cdot n)(q \cdot n^{*})}{q^2}\right] \, , \\
F_{rp}^{*}F_{\mu}^{p} &=& \frac{R^2}{r^2} \eta^{\rho \sigma}
F_{r \rho}^{*}F_{\mu \sigma} = i \frac{R^2}{r^2} f f' q_{\mu}
\left[(n \cdot n^{*}) - \frac{(q\cdot n)(q \cdot n^{*})}{q^2}\right]
\, , \\
F_{\mu p}^{*}F_{\nu}^{p} &=& \frac{R^2}{r^2}\eta^{\rho \sigma}
F_{\mu \rho}^{*} F_{\nu \sigma}
+ \frac{r^2}{R^2} F_{\mu r}^{*} F_{\nu r} \nonumber \\
&=& \frac{R^2}{r^2}f^2 \left[q_{\mu}q_{\nu}(n^{*}\cdot n) + n_{\mu}^{*}n_{\nu}q^2
- q_{\mu}n_{\nu} (q\cdot n^{*}) - q_{\nu}n_{\mu}^{*} (q\cdot n) \right] \\
&& + \frac{r^2}{R^2} (f')^2 \left[n_{\mu}^{*}n_{\nu} + q_{\mu}
q_{\nu} \frac{(q\cdot n)(q \cdot n^{*})}{q^4} - q_{\mu}n_{\nu}
\frac{(q\cdot n^{*})}{q^2} - q_{\nu}n_{\mu}^{*} \frac{(q\cdot
n)}{q^2}  \right] \nonumber \\
&=& \left(n^*_\mu-q_\mu \frac{q \cdot n^*}{q^2}\right) 
\left(n_\nu-q_\nu \frac{q \cdot n}{q^2}\right) \frac{\w^4}{R^2}[K^2_0(\w)+K^2_1(\w)] \nn \\
& & + q_\mu q_\nu \left(n^* \cdot n - \frac{(q \cdot n) (q \cdot n^*)}{q^2}\right) 
\frac{\w^4}{q^2 R^2} K^2_1(\w) \,  , \nn 
\eea
and
\bea
\partial^{r} X^{*}(P)\partial^r X(P) &=&
\frac{r^2\Delta^2}{R^4}|X|^2 , \,\,\,\,\,\,
\partial^{\mu} X^{*}(P)\partial^r X(P) =
-\frac{i\Delta}{r}P^{\mu} |X|^2  , \\
\partial^{\mu} X^{*}(P)\partial^{\nu} X(P) &=& \frac{R^4}{r^4}
P^{\mu}P^{\nu} |X|^2 \, .
\eea
Now, since we want to calculate $T^{\mu \nu}$, and then to contract
it with the leptonic tensor, all terms with factors $q \cdot n$
and/or $q \cdot n^{*}$ will vanish. Therefore, we only need to
expand the effective Lagrangian in terms of the solutions of the
fields. In this way we can express the effective four-point
interaction Lagrangian as
\bea
{\cal {L}}_4^{eff} &=& {\cal {L}}_{4, A}^{eff} + {\cal {L}}_{4,
B}^{eff} + {\cal {L}}_{4, C}^{eff} + {\cal {L}}_{4, D}^{eff} \, ,
\eea
where
\bea
{\cal {L}}_{4, A}^{eff} &=& \partial^{r} X^{*}(P)\partial^r X(P)
F_{rp}^{*}F_{r}^{p} = \frac{\Delta^2}{R^2} (f')^2 (n \cdot n^{*})
|X|^2 \, , \nn \\
{\cal {L}}_{4, B}^{eff}  &=& \partial^{\mu} X^{*}(P)\partial^r X(P)
F_{\mu p}^{*}F_{r}^{p} =\frac{R^2 \Delta}{r^3} f f'(n \cdot n^{*})
|X|^2 (P \cdot q) = {\cal {L}}_{4, C}^{eff} \, , \nn \\
{\cal {L}}_{4, D}^{eff} &=& \partial^{\mu} X^{*}(P)\partial^{\nu}
X(P) F_{\mu p}^{*}F_{\nu}^{p} = \frac{R^4 |X|^2}{r^4 }
\times \left[ \frac{R^2}{r^2}(n \cdot n^{*}) f^2 (P\cdot q)^2
+ \right. \nonumber \\
&& \left.  \left(\frac{R^2}{r^2} q^2 f^2 + \frac{r^2}{R^2}(f')^2\right)
(P\cdot n)(P\cdot n^{*})\right] \, . \nn
\eea
It is worth noting that the terms ${\cal {L}}_{4, A}^{eff}$, ${\cal
{L}}_{4, B}^{eff}$ and ${\cal {L}}_{4, C}^{eff}$ are sub-leading in
comparison with the term in ${\cal {L}}_{4, D}^{eff}$ associated
with $(n \cdot n^{*})$. This term contains a factor $(P \cdot q)^2 \propto s^2 $,
therefore one can neglect the contribution from ${\cal {L}}_{4,
A}^{eff}$, ${\cal {L}}_{4, B}^{eff}$ and ${\cal {L}}_{4, C}^{eff}$
\footnote{This is analogous to what happens to the second and the
third terms in Eq.(82) of reference \cite{Polchinski:2002jw}, which
do not contribute to the hadronic tensor calculation as mentioned in
a previous footnote.}. Now, in order to obtain the imaginary part of
$T^{\mu \nu}$ we have to extract the polarizations $n$ and $n^{*}$.
The factors in front of the tensors $\eta^{\mu \nu}$ and
$P^{\mu}P^{\nu}$ will result in the contributions to the
structure functions $F_1$ and $F_2$, respectively. Thus, we obtain
\bea
\text{Im}_{exc} T^{\mu \nu} &=& \frac{\pi \alpha'}{8} \sum_{m=1}^{\infty} \int d\Omega_3 \, dr \, \sqrt{-g_{ab}} \,
v^i v_i \, \delta\left(m-\frac{\alpha' \tilde s}{4}\right) \times \nn \\
&& \left\{P^{\mu}P^{\nu} \times \left[ \frac{R^4
|X|^2}{r^4} \left(\frac{R^2}{r^2} q^2 f^2 +
\frac{r^2}{R^2}(f')^2\right)\right] + \, \eta^{\mu \nu} \times
\left[ \frac{R^6|X|^2}{r^6}f^2(P\cdot q)^2 \right] \right\}. \nn \\
&&
\label{Tscalar}
\eea
%

\subsection{New relations between $F_1$ and $F_2$ at small $x$}

Our final goal in this section is to calculate the structure functions $F_1$ and
$F_2$ out of comparison of equation (\ref{Tscalar}) with the
Lorentz-tensor decomposition of the spin-zero hadronic tensor
presented in the Introduction. Recall that we use the expression
(\ref{iTmunu}) for the imaginary part of $T_{\mu\nu}$, which has an
integral in the radial coordinate $r$ and in the angular variables
on $S^3$. Thus, we have a normalization condition for the spherical
harmonics on $S^3$ (a similar condition holds for any $S^{p-4}$ in the general
case that we will consider in Section 4)
\bea
\int d\Omega_3 \sqrt{\tilde{g}} v_i v^i Y(\Omega_{3})
Y^{*}(\Omega_{3}) = \rho_3 R^2, \label{angular0}
\eea
as well as the following result for the integral of the square of
the modified Bessel function of the second kind weighted by integer powers of
its argument
\bea
I_{j,n} = \int_{0}^{\infty} d\w \, \w^n \, K_j^2(\w) \, ,
\eea
where $\w = \frac{q R^2}{r}$.
Also, we can define the two following useful integrals
\bea
I_1&=& \frac{\pi \alpha'}{8} \sum_{m=1}^{\infty} R^6
q^2 \int d\Omega_3 dr \sqrt{g_{8}} v_i v^i |X|^2 \frac{f^2(r)}{r^6}
\delta \left(m - \frac{\alpha' s R^2}{4 r^2}\right) \, ,  \label{I1a}
\eea
\bea
I_0&=& \frac{\pi \alpha'}{8} \sum_{m=1}^{\infty} R^2 \int d\Omega_3
dr \sqrt{g_{8}} v_i v^i |X|^2 \frac{(f'(r))^2}{r^2}
\delta \left(m - \frac{\alpha' s R^2}{4 r^2}\right) \, . \label{I2a}
\eea
In order to study $I_1$ we use the following 
definitions $r_0 = \Lambda R^2$, $r_m^2 = \frac{\alpha' s R^2}{4 m}$,
$f(r(\w)) = \w K_1(\w)$ and $f'(r) = \partial_r f(r) =
\frac{\w^3}{qR^2}K_0(\w)$, also Eq.(\ref{angular0}). For $r_m$ and
$\w$ we can replace the sum by an integral in $m$ and we also can
carry out an integral in $\w$ which becomes $\w_m =
\frac{qR^2}{r_m}$. It must be taken into account that
\beq
dm = \frac{\w}{2x\sqrt{4 \pi g_c N}} d\w \, . \nonumber
\eeq
So, the explicit calculation of $I_1$ is as follows
\bea
I_1 &=&  \frac{\pi \alpha'}{8 \Lambda^2} |c_i|^2 \rho_3 R^{5}
 \sum_{m=1}^{\infty} \int dr \left(\frac{r}{R}\right)^3
\left(\frac{r}{r_0}\right)^{-2\Delta} \frac{f^2(r)}{r^6}
\frac{\delta (r-r_m)}{|\frac{\alpha' s R^2}{2 r^3}|} \nn \\
&=& \rho_3 |c_i|^2 \frac{\pi}{4 s \Lambda^2}
\left(\frac{\Lambda^2}{q^2}\right)^{\Delta}
\sum_{m=1}^{\infty} \w^{2 \Delta + 2} K^2_1(\w) \nn \\
&\approx& \rho_3 |c_i|^2 \frac{\pi}{4 s  \Lambda^2}   
\left(\frac{\Lambda^2}{q^2}\right)^{\Delta}
\int_{1}^{\infty}
\frac{d\w}{2x\sqrt{4 \pi g_c N}}\w^{2 \Delta + 3} K^2_1(\w) \nn \\
&=& \rho_3 |c_i|^2 \frac{\pi}{4 s q^2}
\left(\frac{\Lambda^2}{q^2}\right)^{\Delta-1}
\frac{\Lambda^2}{2x\sqrt{4 \pi g_c N}} I_{1,2\Delta+3} \, .
\eea
Similarly, we can calculate the other integral.
Recall that $s= - (P+q)^2 = q^2(-1 + \frac{1}{x} - t_B) \approx \frac{q^2}{x}$.
Thus, we obtain
\bea
I_1 &=& \rho_3 |c_i|^2 \frac{\pi}{8}
\left(\frac{\Lambda^2}{q^2}\right)^{\Delta-1}
\frac{1}{\sqrt{4 \pi g_c N}} I_{1,2\Delta+3} \, , \\
I_0 &=& \rho_3 |c_i|^2 \frac{\pi}{8}
\left(\frac{\Lambda^2}{q^2}\right)^{\Delta-1}
\frac{1}{\sqrt{4 \pi g_c N}} I_{0,2\Delta+3} \, .
\eea
Finally, the comparison with the scalar hadronic tensor given in the
Introduction allows us to extract the structure functions. 
\bea
F_1 &=& \frac{\pi^2}{16 x^2} \rho_3 |c_i|^2 
\left(\frac{\Lambda^2}{q^2}\right)^{\Delta-1}
\frac{1}{\sqrt{4 \pi g_c N}} I_{1,2\Delta+3} \, \\
F_2 &=& \frac{\pi^2}{8 x} \rho_3 |c_i|^2 
\left(\frac{\Lambda^2}{q^2}\right)^{\Delta-1}
\frac{1}{\sqrt{4 \pi g_c N}} (I_{0,2\Delta+3}+I_{1,2\Delta+3}) \, .
\eea
Schematically they are 
$F_1 \propto \frac{1}{x^2} I_1$ and $F_2 \propto \frac{1}{x}(I_1 + I_0)$.
From these functions we straightforwardly obtain a Callan-Gross type relation
\beq 
F_2 = 2 x F_1 \left(1 + \frac{I_0}{I_1}\right) = 2 x F_1
\frac{2\Delta+3}{\Delta+2} \, , 
\eeq 
where we have used the identity
$I_{1,n} = \frac{n+1}{n-1}I_{0,n}$. Notice that the usual
Callan-Gross relation in QCD is $F_2 = 2 x F_1$ while here, like for
glueballs in
\cite{Polchinski:2002jw}, there is an extra factor which only
depends on the scaling dimensions of the scalar mesons
$\Delta=l+2$.

\subsection{Exponentially small $x$ region}

When we considered the parametric region $\exp{(-\sqrt{\lambda})} \ll
x \ll 1/\sqrt{\lambda}$ we dropped the factor ${\tilde s}^{\alpha'
\tilde t/2}$. This factor becomes very important when $x \ll
\exp{(-\sqrt{\lambda})}$. So, let us study the strong coupling
regime $1 \ll \lambda \ll N$, and let us take exponentially large
values of the ten-dimensional Maldestam variable ${\tilde {s}}$ such
that $\frac{\log s}{\sqrt{\lambda}}$ is held fixed.

In this case, the approximations we were considering before break
down because the interaction can no longer be considered local in
the $AdS$ space. What happens in this regime is that the former
parameter $\tilde{t}$ becomes a differential operator. Therefore, it
must be replaced by
\beq
\alpha' \tilde{t} \rightarrow \alpha' \nabla^2 = \alpha' \left(
\frac{R^2t}{r^2} + \nabla^2_{\bot}\right) \label{diffoperator}
\eeq
because we are taking into account the momentum transfer in the
transverse directions \cite{Polchinski:2002jw,Brower:2006ea}.
Consequently, we have to consider a factor
\beq
m^{\alpha'\tilde{t}/2} \sim (\alpha' \tilde{s})^{\alpha'\tilde{t}/2}
\sim x^{-\alpha'\tilde{t}/2} \sim x^{- \alpha' \nabla^2/2}, \label{simeq}
\eeq
within the pre-factor. As in references
\cite{Polchinski:2002jw,Brower:2006ea} the transverse Laplacian is
proportional to $1/R^2$. Thus, the added term is of order
$1/\sqrt{\lambda}$. The idea is that in this parametric regime of
the Bjorken variable a Pomeron is exchanged in the $t$-channel. By
taking into account the transverse momentum transfer it is possible
to obtain the leading correction to the strong coupling limit to the
intercept 2. It turns out that $\tilde s^{\alpha' {\tilde {t}}/2}$ becomes
a diffusion operator. We have to diagonalize the
differential operator (\ref{diffoperator}) in order to obtain the
Regge exponents. In the first place,
the scalar Laplacian operator acting on a field of spin $j$
schematically denoted by $\Phi_j$ is given by
\be
\nabla^2_0 \Phi_j = \frac{1}{\sqrt{-g}} \partial_M[\sqrt{-g}
 g^{MN} \partial_N \Phi_j] \, ,
\ee
where $g_{MN}$ is the ten-dimensional metric. Given the curved
spacetime AdS$_5 \times W^5$, being $W^5$ a compact Einstein
manifold,  the metric is
\be
ds^2 = \frac{r^2}{R^2} \eta_{\mu\nu} dx^\mu dx^\nu + \frac{R^2}{r^2}
dr^2 + R^2 ds^2_W \, .
\ee
Thus, we have the differential operator written as
\be
\alpha' \nabla^2_0 = \alpha' \frac{R^2}{r^2} t + \frac{\alpha'}{R^2}
(r^2 \partial_r^2 + 5 r \partial_r) + \alpha' \nabla^2_W \, .
\ee
Now, let us perform a simple change of variables $e^u = r/r_0$, being
$r_0=\Lambda R^2$ the IR cutoff. The covariant Laplacian operator acting on the
field $\Phi_j$ is defined as
\be
\alpha' {\hat {D}}^2_j \Phi_j = \alpha' e^{2 j u} \nabla^2_0[e^{-2 j
u} \Phi_j] \, .
\ee
Then, in the present case for $j=2$ we obtain the explicit form of
the covariant Laplacian given by
\be
\alpha' {\hat {D}}^2  = \frac{1}{\sqrt{\lambda}} \, t \, e^{-2u} \,
R^2 + \frac{1}{\sqrt{\lambda}} (\partial_u^2 - 4) + \alpha'
\nabla^2_W \, , \\
\ee
where we have dropped the subindex $j=2$ from the operator ${\hat
{D}}^2$. Then, following \cite{Polchinski:2002jw} we modify
Eq.(\ref{iTmunu}) in order to include the effect of the diffusion
operator ${\tilde {s}}^{\alpha' {\tilde {t}}/2} \sim x^{-\alpha'
{\tilde {t}}/2}$. Thus, we have the integral
\bea
&& n^*_{\mu} n_{\nu} Im_{exc}T^{\mu \nu} = \\
&& \frac{\pi \alpha'}{8}
\sum_{m=1}^{\infty} \int d\Omega_3 \, dr \, \sqrt{-g} v_{i}v^{i} \,
F_{kp}^{*}(q)F_{n}^{p}(q) \, \left(\frac{\alpha' \tilde s}{4}\right)^{\alpha' \tilde t/2} \, \left(
\partial^{k} X^{*}(P)\partial^n X(P)\right) \delta\left(m-\frac{\alpha' \tilde {s}}{4}\right) .
\nn 
\label{iTmunusmallx}
\eea
The action of $x^{-\alpha' {\tilde {t}}/2}$ on $\partial^k
X^{*}(P)\partial^n X(P)$ is essentially given by  ${\hat {D}}^2
\left(\partial^{k} X^{*}(P)\partial^n X(P)\right)$ which is
proportional to $P^k P^n {\hat {D}}^2 (X^{*}(P) X(P))$. So, the
exponential of ${\hat {D}}^2$ is non-local, and the diffusion time
is $\ln x$. Now, the point is that when $x$ is exponentially small
the geometric structure corresponding to the region $r_0$ where the
conformal symmetry is broken becomes relevant. In this region we
continue using the same expressions for the field strength because
they are not affected by the conformal symmetry breaking due to
their exponential falloff\footnote{The asymptotic form of
$K_\alpha(y)$ goes like $\sqrt{\pi/(2y)} \exp(-y)$ for $y \gg
|\alpha^2-1/4|$.}. Then, we obtain the expression for the structure functions
\bea
F_1 &=& \frac{\pi^2 \alpha' R^2 q^2}{16 x^2} J_1^{D3D7} \, , \\
F_2 &=& \frac{\pi^2 \alpha' R^2 q^2}{8 x} (J_0^{D3D7}+J_1^{D3D7}) \, , 
\eea
where the integrals are
\bea
J_j^{D3D7} &=& \int d\w \int d\Omega_3 \sqrt{{\hat {g}}_{\Omega_3}} v^i v_i 
\w^3 K_j^2(\w) \left(\frac{\alpha' \tilde s}{4}\right)^{\alpha' \tilde t/2} (X^* X)  \, .
\eea
Now, let us consider the limit for $x$ being exponentially small, then in Eq.(\ref{simeq}) we can
replace the differential operator $-\nabla^2=-D^2$ by its 
the smallest eigenvalue $\zeta$. Therefore, the
expected behaviour for the structure functions is recovered since $F_1 \propto x^{-2+\alpha' \zeta/2}$
and $F_2 \propto x^{-1+\alpha' \zeta/2}$, where the Pomeron intercept is identified as $2-\alpha' \zeta/2$.

Next, in order to solve the integrals $J_j^{D3D7}$ we have to study the falloff of the lowest
eigenfunction of $\nabla_0^2$ which is $r^{-4}$ and then the corresponding one of the lowest
eigenfunction of $\nabla^2$, which is $r^{-2}$. Thus, the integrals are dominated by $\w \approx 1$ leading to 
\bea
J_j^{D3D7} & \propto & \int d\w \int d\Omega_3 \sqrt{{\hat {g}}_{\Omega_3}} v^i v_i 
\w^3 K_j^2(\w) r^{-2}  \, .
\eea
Finally we obtain
\bea
F_1 &=& \frac{\pi^2}{16 x^2} \rho_3 |c_i|^2 
\left(\frac{\Lambda^2}{q^2}\right)^{\Delta-1}
\frac{1}{\sqrt{4 \pi g_c N}} I_{1,2\Delta+3} \,
 x^{\alpha' \zeta/2} \propto x^{-(2+\alpha' |\zeta|/2)} \, , \\
F_2 &=& \frac{\pi^2}{8 x} \rho_3 |c_i|^2 
\left(\frac{\Lambda^2}{q^2}\right)^{\Delta-1}
\frac{1}{\sqrt{4 \pi g_c N}} (I_{0,2\Delta+3}+I_{1,2\Delta+3})  \, x^{\alpha' \zeta/2} 
\propto x^{-(1+\alpha' |\zeta|/2)} \, ,
\eea
where the eigenvalue of the diffusion operator is $\zeta = \frac{4}{R^2}(1-\Delta^2)\leq 0$.
In addition, the Callan-Gross type relation is the same as when 
$\exp{(-\sqrt{\lambda})} \ll x \ll 1/\sqrt{\lambda}$
\beq 
F_2 = 2 x F_1 \left(1 + \frac{I_0}{I_1}\right) = 2 x F_1
\frac{2\Delta+3}{\Delta+2} \, . \nn 
\eeq 

In the case of vector mesons the calculations are similar to those presented here for
scalar mesons. The Callan-Gross type relations are the same as the ones obtained in the regime
where $\exp{(-\sqrt{\lambda})} \ll x \ll 1/\sqrt{\lambda}$, 
which are discussed in Section 3. The difference now 
is that the eigenvalue of the diffusion operator changes.

\section{DIS from vector mesons at small $x$ from string theory}

In the low-energy regime parallel fluctuations on a
single flavor Dp-brane world-volume are described by a $U(1)$
gauge field $B_a$. This gauge field is interpreted as the
holographic dual description of a vector meson composed by a quark
and an anti-quark with $U(1)$ flavor symmetry, {\it i.e.} the $U(1)$
gauge symmetry on the Dp-brane world-volume becomes a $U(1)$ global
(flavor) symmetry in the dual gauge theory description. In string
theory a dynamical meson is described by an oriented open string
whose endpoints are attached to a flavor Dp-brane.

We are interested in obtaining the hadronic tensor for polarized
vector mesons. It is easy to check that the hadronic tensor
(\ref{DIS16}) can be split into a symmetric plus an anti-symmetric
parts as follows
\begin{equation}\label{DISWsa}
W_{\mu\nu}(P,q)_{h'\,h}^{vector} = W^S_{\mu\nu}(P,q)_{h'\,h} +
W^A_{\mu\nu}(P,q)_{h'\,h}  \, .
\end{equation}
In the limit $|t|\ll 1$ ($\tilde\kappa\simeq 1$) these contributions
can be written as
\bea
W_{\mu\nu}^S&=&\left\{F_1+\left[\frac{(q\cdot\zeta^*)(q\cdot\zeta)}{(P\cdot
q)^2}
-\frac{1}{3}\right]b_1 +\frac{P^2}{9(P\cdot q)}(-b_2+3b_3)\right\}\eta_{\mu\nu}\\
&&-\Bigg\{3F_2+\frac{(q\cdot\zeta)(q\cdot\zeta^*)}{(P\cdot
q)^2}(b_2+3b_3+3b_4)-b_2
\Bigg\}\frac{P_\mu P_\nu}{3(P\cdot q)}\nn\\
&&+(-b_2+3b_3)\frac{(\zeta_\mu\zeta_\nu^*+\zeta_\mu^*\zeta_\nu)}{6(P\cdot
q)} +\frac{(-b_2+3b_4)}{12(P\cdot q)^2}\left[(q\cdot
\zeta^*)(P_\mu\zeta_\nu+\zeta_\mu P_\nu)+c.c.\right] ,
\label{decompositionsim}  \nn \\
W_{\mu\nu}^A&=&\left[g_1+\left(2-\frac{q^2P^2}{(P\cdot
q)^2}\right)g_2\right]
\frac{1}{P^2}(\zeta_{\nu}\zeta_{\mu}^{*}-\zeta_{\mu}\zeta_{\nu}^{*}) \nn\\
&&+\frac{(g_1+2g_2)}{(P\cdot q)P^2}\left[(q\cdot
\zeta^*)(P_{\mu}\zeta_{\nu} - P_{\nu}
\zeta_{\mu})+c.c.\right] \, ,\label{decompositionantisim}
\eea
where {\it c.c.} denotes complex conjugate.

In order to construct the hadronic tensor for vector mesons we
consider the four-point scattering amplitude of two open and two
closed strings ${\cal {A}}_4^{2o2c, vector}$. As before the closed
strings which are gravitons in a particular polarization state
represent two virtual photons, while the open strings represent
vector mesons on the flavor Dp-brane. The calculation we describe
below holds for both type IIA and type IIB string theory Dp-brane
models. However, in order to give a concrete example we first
consider the D3D7-brane system. Then, in Section 4 we derive a more
general expression which also holds for the
D4D6$\mathrm{\overline{D6}}$-brane and
D4D8$\mathrm{\overline{D8}}$-brane systems. Notice that although in
this section we consider vector mesons, we also recover the results
for scalar mesons presented in Section 2, as a particular simpler
case of this formalism, which is also a consistency check of our
calculations.

\subsection{Four-point graviton-vector meson tree-level amplitudes}

Let us consider the four-point scattering amplitude of two open and
two closed strings ${\cal {A}}_4^{2o2c, vector}$. We consider the
leading contribution, {\it i.e.} genus zero surface, which
corresponds to the planar limit in the dual gauge theory. We may
split ${\cal {A}}_4^{2o2c, vector}$ as sum of several contributions
\cite{Stieberger:2009hq}
\be
{\cal {A}}_4^{2o2c, vector} =  {\cal {P}}_1^{2o2c, vector} \, {\cal
{K}}_1^{2o2c, vector} + {\cal {P}}_2^{2o2c, vector} \, {\cal
{K}}_2^{2o2c, vector} + \cdot \cdot \cdot \, .
\ee
As in the previous section we can obtain an effective
four-point interaction Lagrangian involving two vector mesons ${\cal
{L}}_{hhBB}^{eff}$, by evaluating ${\cal {K}}_1^{2o2c,
vector}|_{\tilde t=0}$.  This effective Lagrangian describes the
interaction of two gravitons in certain polarization states and two
$B_a$ fields (representing the holographic dual of two vector
mesons).

Alternatively, we can proceed in a different way. One may consider
the low-energy limit of the full string theory action, containing
open strings attached to Dp-branes, closed strings in the
background, and open-closed strings interactions,
by considering the $\alpha' \rightarrow 0$
limit. This leads to the ten-dimensional supergravity action,
$S_{sugra}$, which contains gravitons (and also the dilaton and the
antisymmetric tensor field) and the Dirac-Born-Infeld action as the
low-energy limit of the flavor brane, $S_{DBI}$. On the one hand, from $S_{sugra}$ we
obtain the graviton propagator and the graviton interaction
vertices. In addition, $S_{DBI}$ leads to the $B_a$ gauge
field propagator on the brane world-volume. Also, from $S_{DBI}$ we
extract a four-point interaction term $L_{hhBB}$ and a three-point
interaction term $L_{hBB}$, which lead to the corresponding
interaction vertices. Now, with these building blocks we can draw
all the Feynman diagrams which contribute at tree level: the
four-point contact term, and the $s$-, $u$- and $t$-channels,
respectively. The graviton pole of the $t$-channel gives the
kinematic part of the effective Lagrangian ${\cal {L}}_{hhBB}^{eff}$ mentioned above. In
the present case recall that two limits have been taken, $\alpha'
\rightarrow 0$ and then ${\tilde {t}} \rightarrow 0$, which in addition to the pre-factor render
${\cal {L}}_{hhBB}^{eff}$. This pre-factor has been derived in
\cite{Stieberger:2009hq} in terms of integrals in the world-sheet.
We can assume that it is similar to the
corresponding pre-factor obtained for scalar mesons. We have done
the calculations of the hadronic tensor under this assumption,
obtaining results which are consistent with those obtained for
glueballs in \cite{Polchinski:2002jw}. Although this assumption
seems to be a caveat of the calculations presented here, we think
that the eight structure functions and the relations among them that
we have obtained provide strong evidence in favor of its validity.
We plan to further investigate this assumption in a future work.

\subsection{Four-point graviton-vector meson tree-level amplitudes
from supergravity}

Let us start from the Dirac-Born-Infeld action of a single flavor
D7-brane as in Section 2.2
\beq
S_{DBI}=-T_7 \int d^8 \xi \sqrt{- det\left(\hat{P}[g]_{ab} + (2\pi
\alpha')F_{ab}\right)} \, . \nn
\eeq
One can expand the squared root in the DBI action at third order
obtaining the same equation (\ref{taylor}), which we repeat here to
make easier the reading
\beq
\sqrt{- g} \approx \sqrt{-g_0} \left[ 1 + \frac{1}{2} H -
\frac{1}{4}H^a_b H^b_a + \frac{1}{8} H^2 + \frac{1}{6}H^a_b H^b_c
H^c_a - \frac{1}{8} H H^a_b H^b_a + \frac{1}{48} H^3\right] \, , \nn
%
\eeq
where $H \equiv H^a_a$ is the trace and $g_0$ is the unperturbed
background metric. The next order in the expansion above is given by
the following set of terms
\beq
\sqrt{-g_0}\left[\frac{1}{384}H^4 + \frac{1}{32}H^a_b H^b_a (H^c_d
H^d_c - H^2) + \frac{1}{12} H H^a_b H^b_c H^c_a - \frac{1}{8} H^a_b
H^b_c H^c_d H^d_a \right] \, . \label{taylor4}
\eeq
These quartic interaction terms allow one to obtain a four-point
interaction term in the low-energy Lagrangian. In order to do it let
us consider the case where $(g_0)_{ab} = \eta_{ab}$. If we consider the
static gauge and choose the graviton polarizations to be parallel to
the D7-brane, we can set
\beq
H_{ab} \rightarrow 2\kappa h_{ab} + (2 \pi \alpha') G_{ab} \, ,
\eeq
where the $U(1)$ field strength on the D7-brane is $G_{ab}=
\partial_a B_b - \partial_b B_a$.\footnote{Notice that the field strength $F_{\mu\nu}=
\partial_\mu A_\nu - \partial_\nu A_\mu$ corresponds to the on-shell graviton, having
used the {\it Ansatz} $h \sim A v$.}

Now, we can straightforwardly extract the Lagrangian term associated
with the quartic interaction $hhBB$ (shown in figure 5 
along with the rest of the vector meson diagrams) which reads as
\beq
L_{hhBB} = T_7 \kappa^2 \left[\frac{1}{8} h^2 G^{ab}G_{ab} + 2 h h^{ab}
G_{bc} G^c_a - \frac{1}{4} h^{ab}h_{ab} G^{cd} G_{cd} - h^{ab}h^{cd}
G_{bc} G_{da} - 2 h^{ab}h_{bc} G^{cd} G_{da} \right].
\eeq
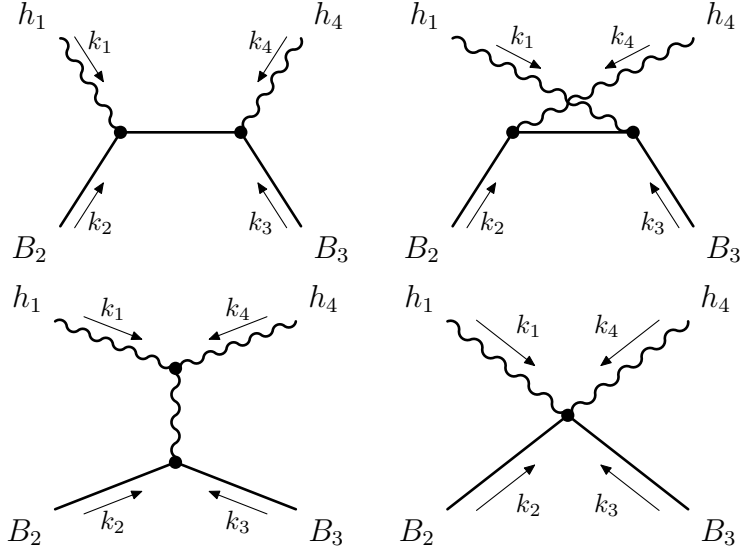
\begin{figure}
\unitlength=1mm
\begin{center}
\begin{fmffile}{Schannel} 
\begin{fmfgraph*}(40,25)
\fmfleft{i1,i2}
\fmfright{o1,o2}
\fmf{plain}{i1,v1}
\fmf{photon}{i2,v1}
\fmf{plain}{v2,o1}
\fmf{photon}{v2,o2}
\fmf{plain}{v1,v2}
\fmfdot{v1,v2}
\fmflabel{$h_1$}{i2}
\fmflabel{$h_4$}{o2}
\fmflabel{$B_2$}{i1}
\fmflabel{$B_3$}{o1}
			 \marrowup{a}{up}{top}{$k_1$}{i2,v1}
       \marrowup{b}{down}{bot}{$k_2$}{i1,v1}
       \marrowup{d}{up}{top}{$k_4$}{o2,v2}
       \marrowup{e}{down}{bot}{$k_3$}{o1,v2}
\end{fmfgraph*}
\end{fmffile}
\hspace{15pt} 
\begin{fmffile}{Uchannel}
\begin{fmfgraph*}(40,25)
\fmfleft{i1,i2}
\fmfright{o1,o2}
\fmf{phantom}{i2,v1}
\fmf{phantom}{v2,o2}
\fmf{plain}{v1,i1}
\fmf{plain}{v2,o1}
\fmf{plain}{v1,v2}
\fmf{photon,tension=0}{v1,o2}
\fmf{photon,tension=0}{v2,i2}
\fmfdot{v1,v2}
\fmflabel{$h_1$}{i2}
\fmflabel{$h_4$}{o2}
\fmflabel{$B_2$}{i1}
\fmflabel{$B_3$}{o1}
			 \marrowshort{a}{up}{top}{$k_1$}{i2,v2}
       \marrowup{b}{down}{bot}{$k_2$}{i1,v1}
       \marrowshort{d}{up}{top}{$k_4$}{o2,v1}
       \marrowup{e}{down}{bot}{$k_3$}{o1,v2}
\end{fmfgraph*}
\end{fmffile}

\vspace{35pt}

\begin{fmffile}{Tchannel}
\begin{fmfgraph*}(40,25)
\fmfleft{i1,i2}
\fmfright{o1,o2}
\fmf{photon}{i2,v1,o2}
\fmf{plain}{i1,v2,o1}
\fmf{photon}{v1,v2}
\fmfdot{v1,v2}
\fmflabel{$h_1$}{i2}
\fmflabel{$h_4$}{o2}
\fmflabel{$B_2$}{i1}
\fmflabel{$B_3$}{o1}
      \marrow{a}{up}{top}{$k_1$}{i2,v1}
       \marrow{b}{up}{top}{$k_4$}{o2,v1}
       \marrow{d}{down}{bot}{$k_3$}{o1,v2}
       \marrow{e}{down}{bot}{$k_2$}{i1,v2}
\end{fmfgraph*}
\end{fmffile}
\hspace{15pt}
\begin{fmffile}{Contact} 
\begin{fmfgraph*}(40,25)
\fmfleft{i1,i2}
\fmfright{o1,o2}
\fmf{photon}{i2,v1,o2}
\fmf{plain}{i1,v1,o1}
\fmfdot{v1}
\fmflabel{$h_1$}{i2}
\fmflabel{$h_4$}{o2}
\fmflabel{$B_2$}{i1}
\fmflabel{$B_3$}{o1}
			 \marrowup{a}{up}{top}{$k_1$}{i2,v1}
       \marrowup{b}{up}{top}{$k_4$}{o2,v1}
       \marrowup{d}{down}{bot}{$k_3$}{o1,v1}
       \marrowup{e}{down}{bot}{$k_2$}{i1,v1}
\end{fmfgraph*}
\end{fmffile}
\vspace{10pt}
{\caption{\small The four Feynman diagram corresponding to the
holographic calculation of the four-point amplitude in the vector meson case. 
The later are represented by solid lines.}} \label{FigContactVec}
\end{center}
\end{figure}
\unitlength=1pt
Notice that the first two terms in $L_{hhBB}$ will not
contribute to the amplitude because the graviton polarization is
traceless. We can show that the contact diagram gives an amplitude
of the form
\bea
&& A_{4}^{vector}(h_1,h_4,\e_2,\e_3) =  \\
&&- i T_7 \kappa^2 [2 (h_1 h_4)[(\e_2 k_3)(\e_3 k_2)+(k_2 k_3)(\e_2 \e_3)]  \nonumber \\
&& + 4 [(k_2 h_1 k_3)(\e_2 h_4 \e_3) + (\e_2 h_1 \e_3)(k_2 h_4 k_3)
- (\e_2 h_1 k_3)(\e_3 h_4 k_2) - (\e_3 h_1 k_2)(\e_2 h_4 k_3)] \nonumber \\
&& + 4 [(k_2 k_3)(\e_2 h_1 h_4 \e_3) + (k_2 k_3)(\e_3 h_1 h_4 \e_2)
+ (\e_2 \e_3) (k_2 h_1 h_4 k_3) + (\e_2 \e_3) (k_3 h_1 h_4 k_2)]       \nonumber \\
&&  -4 [(\e_2 k_3 )(\e_3 h_1 h_4 k_2) (\e_2 k_3 )(k_2 h_1 h_4 \e_3)
+ (\e_3 k_2 )(\e_2 h_1 h_4 k_3) + (\e_3 k_2 )(k_3 h_1 h_4 \e_2)] ] .
\nonumber 
\eea
There are no terms like $(\e_2 h_1 k_2)$ since they can only come
from the term $h^{ab} G_{ab}$ in $L_{hhBB}$. However, this term
trivially vanishes because the graviton is a symmetric tensor and
the field strength is an antisymmetric one. Thus, we have obtained a
contact term with two external gravitons and two external vector
mesons.

\subsection{The $s$- and $u$-channels}

In order to calculate the Feynman diagrams corresponding to the $s$-
and $u$-channels indicated in figure 5, we only need to know
the interaction vertex between a graviton and two gauge fields on
the D7-brane, $L_{hBB}$, and the gauge field propagator, which we
obtain from the $U(1)$ gauge field kinetic term $L_{BB}$. Both are
computed by using similar methods as described in the previous
section. Thus, we obtain
\bea
L_{BB} = -\frac{T_7}{4} G^{ab} G_{ab} \, \ , \ \, L_{hBB} = T_7 \kappa
\left[\frac{1}{4} h G^{ab} G_{ab} + h^{ab} G_{bc} G^c_a\right]
\label{LhBB}  \, .
\eea
From the kinetic term for the $U(1)$ gauge field $B_a$, the gauge
field propagator is $D_{ab}(k) = -i \eta_{ab}/(T_7 k^2)$. On the
other hand, since there are on-shell gravitons the three-point
interaction vertex can be derived from $L_{hBB}$, obtaining
\beq
V_{hBB}^{ab, c, d}(k_2,k_3) = 2 i \kappa T_7 \left[\eta^{db} k_3^{c} k_2^{a}  + \eta^{cb}
k_2^{d} k_3^{a}  - \eta^{cd} k_2^{a} k_3^{b}  -
(k_2 \cdot k_3) \eta^{ca} \eta^{db}\right] \, ,
\eeq
where $k_2$ and $k_3$ are the momenta of the incoming vector mesons
on the D7-brane world-volume.
By joining the $V_{hBB}$ vertex with a gauge field (vector meson)
propagator one constructs the $s$-channel Feynman diagram in the
low-energy theory. The $s$-channel four-point amplitude displayed in
figure 5 is given by
\bea
&&A_{s}(h_1,h_4,\e_2,\e_3) = h_{1 gk}
h_{4 cd} \epsilon_{2e} \epsilon_{3f}
V_{hAA}^{gk, e, a}(k_2,-k_1-k_2)
D_{ab}(k_1+k_2) V_{hAA}^{cd ,b, f}(k_1+k_2,k_3) \nonumber \\
&&=- \frac{4i\kappa^2}{s} \left[ -\frac{1}{2} \left[ t (k_2 h_1 \e_2)(k_3
h_4 \e_3) + s [(k_2 h_1 k_2)(\e_2 h_4 \e_3) -  (k_3 h_1 \e_2) (k_3
h_4 \e_3) \right.\right.\nonumber \\
&& + (\e_2 h_1 \e_3)(k_3 h_4 k_3)]]
%
+ \frac{1}{2} s [(\e_2 k_1)(k_2 h_1h_4 \e_3) + (\e_3 k_4)(\e_2
h_1h_4 k_3)]
+ (\e_2 k_1)(\e_3 k_4)(k_2 h_1 h_4 k_3) \nonumber \\
&& + (k_2 h_1 \e_2)[(\e_3 k_4)(k_2 h_4 k_3) - (\e_3 k_2) (k_3 h_4
k_3)]
+ (\e_2 k_1)(k_2 h_1 k_3) [(k_3 h_4 \e_3) - (k_3 h_4 k_3)]\nonumber \\
&& \left. + (k_2 h_1 k_2) [(\e_2 \e_3)(k_3 h_4 k_3) - (\e_2 k_3)(k_3
h_4 \e_3) - (\e_3 k_4)(k_3 h_4 \e_2)] + \frac{1}{4} s^2 (\e_2 h_1
h_4 \e_3) \right] \, .
\eea
The $u$-channel four-point amplitude is straightforwardly obtained
from $A_{s}(h_1,h_4,\e_2,\e_3)$ as indicated below
\ber
A_{u}(h_1,h_4,\e_2,\e_3) = A_s (2 \leftrightarrow 3, u
\leftrightarrow s) \, .
\eer
%

\subsection{The $t$-channel}

Let us now focus on the $t$-channel. We need three pieces: the
three-graviton vertex derived from $L_{hhh}$, the graviton
propagator which is derived from $L^{\partial^2}_{hh}$, and the
three-point interaction vertex with one graviton and two gauge
fields derived from $L_{hBB}$ which we have obtained in the previous
section. Since the graviton connecting the three-graviton vertex from
$L_{hhh}$ and the interaction vertex from $L_{hBB}$ is an off-shell
one, we cannot neglect the first term. Thus, we obtain the
following graviton-vector meson vertex
\bea
&& V_{hBB}^{ab, c , d} (k_1,k_2) = 2 i T_7 \kappa \times
\nonumber \\
&& \left[\eta^{db} k_2^{c} k_1^{a}  + \eta^{cb}
k_1^{d}k_2^{a}  - \eta^{cd}k_1^{a}k_2^{b}  -
(k_1 \cdot k_2) \eta^{ca} \eta^{db} + \frac{1}{2}
\eta^{ab}(\eta^{cd}(k_1 \cdot k_2) - k_1^{c}
k_2^{d}) \right]. \nonumber
\eea
This vertex must then be contracted with Eq.(\ref{eq-sanann}).
Now, we just contract this result with the vertex $V_{hBB}^{ab , c ,
d}$, in order to obtain the $t$-channel amplitude
\bea
A_t = h_{1 gk} h_{4 cd} S^{gk, cd}_{\phantom{gk, cd} ab}(k_1,k_4) V_{hBB}^{ab , e , f}
(k_2,k_3) \epsilon_{2e} \epsilon_{3f} \, .
\eea
Next, let us briefly comment on the two gravitons and two vector
mesons tree-level scattering amplitude. First, notice that $A_t$
gives the following contribution to the two gravitons and two vector
mesons string theory scattering amplitude:
\bea
&& A_t =  \frac{4 i \kappa^2 T_7}{t} \times \frac{1}{2} (h_1 h_4) \left[ (s+\frac{1}{2}u)[(\e_2 k_3)(\e_3
k_1) + (\e_2k_4)(\e_3k_2)] \right. \nn \\
&&  + (u + \frac{1}{2}s) [(\e_2k_3)(\e_3k_4)
+ (\e_2k_1)(\e_3k_2)] - \frac{1}{2}t[2(\e_2k_1)(\e_3k_1) + 2(\e_2k_4)(\e_3k_4)\nonumber \\
&& +(\e_2 k_4)(\e_3 k_1) + (\e_2k_1)(\e_3k_4) + 3(\e_2k_3)(\e_3k_2)]
 \left. - \frac{1}{2}(\e_2 \e_3) [u^2 + s^2 + su]\right] \nonumber\\
&&
+ (\e_2 \e_3) \left[ (k_2 h_1 k_3)(k_1h_4k_1) + (k_4 h_1 k_4)(k_2h_4k_3)
 - (k_4h_1k_3)(k_2h_4k_1) \right. \nonumber \\
 &&\left. - (k_4h_1k_2)(k_3h_4k_1) + \frac{1}{2}u[(k_2 h_1 h_4 k_1)
 + (k_4 h_1 h_4 k_3)] + \frac{1}{2}s[(k_3 h_1 h_4 k_1) + (k_4 h_1 h_4 k_2)] \right. \nonumber\\
 &&\left.- \frac{1}{2}t[(k_3 h_1 h_4 k_2)+ (k_2 h_1 h_4 k_3)+ (k_4 h_1 h_4 k_1)]\right]
 - \frac{1}{2} t \left[(\e_2 h_1 \e_3)(k_1 h_4 k_1) + (k_4 h_1 k_4)(\e_2 h_4 \e_3)\right] \nonumber\\
 && + \frac{1}{4}t^2 \left[(\e_2 h_1 h_4 \e_3) + (\e_3 h_1 h_4 \e_2)\right] + \frac{1}{2} (\e_2 k_3)\left[ -2 (k_4 h_1 k_4)(\e_3 h_4 k_2) -2 (\e_3 h_1 k_2)(k_1 h_4 k_1)  \right. \nonumber\\
&& + 2(k_4 h_1 k_2)(\e_3 h_4 k_1) + 2 (\e_3 h_1 k_4)(k_1 h_4 k_2)  + t (k_2h_1h_4\e_3) + t (\e_3 h_1 h_4 k_2) - s(\e_3 h_1 h_4 k_1)   \nonumber\\
&&- u (k_4 h_1 h_4 \e_3)] + \frac{1}{2} (\e_3 k_2) \left[ -2 (k_4 h_1 k_4)(\e_2 h_4 k_3)  -2 (\e_2 h_1 k_3)(k_1 h_4 k_1) + 2(k_4 h_1 k_3)(\e_2 h_4 k_1)  \right.\nonumber\\
&& \left.+ 2 (\e_2 h_1 k_4)(k_1 h_4 k_3) + t (k_3h_1h_4\e_2) + t (\e_2 h_1 h_4 k_3) - u(\e_2 h_1 h_4 k_1) - s (k_4 h_1 h_4 \e_2)\right]
  \nonumber\\
&& + \frac{1}{2} [ (\e_2 h_1 h_4 k_1)(\e_3 k_1)  + (\e_3 h_1 h_4 k_1)(\e_2 k_1) + (k_4 h_1 h_4 \e_2)(\e_3k_4) + (k_4 h_1 h_4 \e_3)(\e_2k_4) \nonumber\\
&&  + (\e_2 h_1 k_4)(\e_3 h_4 k_1) + (\e_3 h_1 k_4)(\e_2 h_4 k_1)] + \left[(k_3 h_1 h_4 k_1)(\e_2 k_4)(\e_3 k_2)
 +(k_2 h_1 h_4 k_1)(\e_2 k_3)(\e_3 k_4) \right.\nonumber\\
&& \left. + (k_2 h_4 h_1 k_4)(\e_2 k_3)(\e_3 k_1) + (k_3 h_4 h_1 k_4)(\e_2 k_1)(\e_3 k_2) - (k_4 h_1 h_4 k_1)(\e_2 k_3)(\e_3 k_2)\right]
\eea
This expression can be checked by noting that if we only keep terms
which have the factor $(\e_2\e_3)$, we should recover the previous
results for scalar mesons. A similar comment also holds for the
other tree-level amplitudes $A_c$, $A_s$ and $A_u$. This is because
the string theory calculation turns out to be the same, recall that
in the case of scalar mesons the polarization is perpendicular to
the brane world-volume coordinates. Therefore, the rest of momenta
and the graviton polarizations can only be contracted among
themselves.

\subsection{Effective four-point interaction Lagrangian for
vector mesons}

Now, we proceed in an analogous way as we have done for the scalar
mesons. The difference is that the calculations are much more
involved. We first list the terms which are necessary to reproduce
the coefficient of the $t$-channel graviton pole. The process
involves four external fields, two gravitons and two vector mesons.
Recall that the vector mesons are massless modes of open strings,
and their polarizations are parallel to the flavor-brane
world-volume. There are two types of such terms. On the one hand,
there are terms similar to the case of the scalar mesons, so we
already know their amplitudes, and we will not repeat them here. The
only difference from these terms in comparison with the scalar case
is that there is an extra factor of the form $(\e_2 \e_3)$. On the
other hand, there is also a new type of terms, for which in the
first and second part of the Lagrangian they share $0$, $2$ or $4$
indices. These terms are \footnote{This list does not
include all possible terms but only those which are necessary for
the construction of an effective four-point interaction Lagrangian
for vector mesons describing the $t$-channel amplitude.}

\bea
{\cal {L}}_{1} = \partial_{a} B_{b} \partial^{b} B^{a}
\partial^{d} h^{ef} \partial_{d} h_{ef} &\rightarrow& -2 t (h_1 h_4) (\e_2 k_3)(\e_3 k_2)   \nonumber \\
{\cal {L}}_{2} = \partial_{a} B_{b} \partial^{b} B^{a}
\partial^{d} h_{ef} \partial_{f} h_{de} &\rightarrow& 4 (\e_2 k_3)(\e_3 k_2) (k_4 h_1 h_4 k_1) \nonumber\\
{\cal {L}}_{3} = \partial_{a} B^{d} \partial^{a} B^{e}
\partial_{d} h^{cf} \partial_{e} h_{cf} &\rightarrow&
- t (h_1 h_4) [(\e_2 k_1)(\e_3 k_4) + (\e_2 k_4)(\e_3 k_1)]\nonumber\\
{\cal {L}}_{4} = \partial_{a} B^{d} \partial^{a} B^{e}
\partial_{d} h^{cf} \partial_{c} h_{ef} &\rightarrow&
- \frac{1}{2} t \left[(\e_2 k_1)(k_4 h_1 h_4 \e_3)
+ (\e_3 k_1)(k_4 h_1 h_4 \e_2) \right. \nonumber \\
&&\left. + (\e_2 k_4)(\e_3 h_1 h_4 k_1) + (\e_3 k_4)(\e_2 h_1 h_4 k_1)\right] \nonumber\\
{\cal {L}}_{5} = \partial_{a} B^{d} \partial^{a} B^{e}
\partial^{c} h_{df} \partial_{c} h_{e}^{f} &\rightarrow&
\frac{1}{2} t^2 [(\e_2 h_1 h_4 \e_3) + (\e_3 h_1 h_4 \e_2)] \nonumber\\
{\cal {L}}_{6} = \partial_{a} B^{d} \partial^{a} B^{e}
\partial^{c} h_{df} \partial^{f} h_{ec} &\rightarrow&
-t [(\e_2 h_1 k_4)(\e_3 h_4 k_1) + (\e_3 h_1 k_4)(\e_2 h_4 k_1)]\nonumber\\
{\cal {L}}_{7} = \partial^{d} B_{a} \partial^{a} B^{e}
\partial_{d} h^{cf} \partial_{e} h_{cf} &\rightarrow&
\frac{1}{2}(h_1h_4) \left[-s((\e_2 k_1)(\e_3 k_2) +(\e_2 k_3)(\e_3 k_4)) \right. \nonumber \\
&& \left. - u ((\e_2 k_3)(\e_3 k_1)+(\e_2 k_4)(\e_3 k_2))\right] \nonumber\\
{\cal {L}}_{8} = \partial^{d} B_{a} \partial^{a} B^{e}
\partial^{c} h_{df} \partial_{c} h_{e}^{f} &\rightarrow&
-\frac{1}{2}t \left[(\e_2 k_3)((\e_3 h_1 h_4 k_2)+(k_2 h_1 h_4 \e_3)) \right. \nonumber \\
&&\left. + (\e_3k_2)((\e_2 h_1 h_4 k_3)+(k_3 h_1 h_4 \e_2))\right]\nonumber\\
{\cal {L}}_{9} = \partial^{d} B_{a} \partial^{a} B^{e}
\partial^{c} h_{df} \partial^{f} h_{ec} &\rightarrow&
(\e_2 k_3)[(\e_3 h_1 k_4)(k_2 h_4 k_1) +(k_2 h_1 k_4)(\e_3 h_4 k_1)] \nonumber \\
&&+ (\e_3 k_2)[(\e_2 h_1 k_4)(k_3 h_4 k_1) +(k_3 h_1 k_4)(\e_2 h_4 k_1)]  \nonumber\\
{\cal {L}}_{10} = \partial^{d} B_{a} \partial^{a} B^{e}
\partial_{d} h^{cf} \partial_{c} h_{ef} &\rightarrow& - \frac{1}{2}s [(\e_2 k_3)(k_4 h_1 h_4 \e_3)
+ (\e_3 k_2)(\e_2 h_1 h_4 k_1)] \nonumber \\ &&- \frac{1}{2}u [(\e_2
k_3)(\e_3 h_1 h_4 k_1)
+ (\e_3 k_2)(k_4 h_1 h_4 \e_2)]  \nonumber\\
{\cal {L}}_{11} = \partial^{d} B_{a} \partial^{a} B^{e}
\partial_{e} h^{cf} \partial_{c} h_{df} &\rightarrow&
(\e_2 k_3)[(\e_3 k_1)(k_4 h_1 h_4 k_2) +(\e_3 k_4)(k_2 h_1 h_4 k_1)] \nonumber \\
&& + (\e_3 k_2)[(\e_2 k_1)(k_4 h_1 h_4 k_3) + (\e_2 k_4)(k_3 h_1 h_4 k_1)] \nonumber\\
{\cal {L}}_{12} = \partial_{a} B^{d} \partial^{a} B^{e} h^{fg}
\partial_{f} \partial_{g} h_{de} &\rightarrow&
-t [(\e_2 h_1 \e_3) (k_1 h_4 k_1) + (k_4 h_1 k_4)(\e_2 h_4 \e_3)]\nonumber \\
{\cal {L}}_{13} = \partial^{d} B_{a} \partial^{a} B^{e} h^{fg}
\partial_{f} \partial_{g} h_{de} &\rightarrow& (\e_2 k_3)[(k_4 h_1 k_4)
(\e_3 h_4 k_2) + (\e_3 h_1 k_2) (k_1 h_4 k_1)] \nonumber \\
&& + (\e_3 k_2) [(k_4 h_1 k_4) (\e_2 h_4 k_3)
+ (\e_2 h_1 k_3) (k_1 h_4 k_1)] \nonumber \\
{\cal {L}}_{14} = \partial_{a} B^{d} \partial^{a} B^{e} h^{fg}
\partial_{d} \partial_{e} h_{fg} &\rightarrow& -t (h_1 h_4) [(\e_2
k_1)(\e_3 k_1) + (\e_2 k_4)(\e_3 k_4)] \, . \nonumber
\eea
Now, by taking into account these terms and using four-momentum
conservation and on-shell transversal propagation, the effective
Lagrangian associated to the coefficient of the $t$-channel
amplitude up to a global factor can be written as \footnote{Notice
that due to momentum conservation this form
is not unique. However all these possible effective Lagrangians are
equivalent.}
\bea
-\frac{1}{4} \times {\cal {L}}_{1} - \frac{1}{4}\times {\cal
{L}}_{2} + \frac{1}{4}\times {\cal {L}}_{3} + 1\times {\cal {L}}_{4}
+ \frac{1}{2}\times {\cal {L}}_{5} - \frac{1}{2}\times {\cal
{L}}_{6} + \frac{1}{2}\times {\cal {L}}_{7} - 1\times {\cal {L}}_{8}
+ 1 \times {\cal {L}}_{9} \label{Leff vec} \nonumber \\
 - 1 \times {\cal {L}}_{10} + 1 \times {\cal {L}}_{11}
 + \frac{1}{2} \times {\cal {L}}_{12} - 1\times {\cal {L}}_{13}
 + \frac{1}{2}\times {\cal {L}}_{14} \, . \nonumber
 \label{VectorLagrangian}
\eea
This expression does not include the part analogous to the scalar
meson, neither the terms associated with $(k_1 h_4 k_1)$ and $(k_4
h_1 k_4)$ factors which vanish under the condition that the graviton
is $h^{m i}\rightarrow A^m v^i$. The effective four-point interaction 
is depicted in figure 6.
\begin{figure}
\unitlength=1mm
\begin{center}
\begin{fmffile}{Eff}
\begin{fmfgraph*}(45,30)
\fmfcmd{
    path quadrant, q[], otimes;
    quadrant = (0, 0) -- (0.5, 0) & quartercircle & (0, 0.5) -- (0, 0);
    for i=1 upto 4: q[i] = quadrant rotated (45 + 90*i); endfor
    otimes = q[1] & q[2] & q[3] & q[4] -- cycle;
}
\fmfwizard
\fmfleft{i1,i2}
\fmfright{o1,o2}
\fmf{photon}{i2,v1,o2}
\fmf{plain}{i1,v1,o1}
\fmfv{d.sh=otimes,d.f=empty,d.si=.1w}{v1}
\fmflabel{$h_1$}{i2}
\fmflabel{$h_4$}{o2}
\fmflabel{$B_2$}{i1}
\fmflabel{$B_3$}{o1}
       \marrowup{a}{up}{top}{$k_1$}{i2,v1}
       \marrowup{b}{up}{top}{$k_4$}{o2,v1}
       \marrowup{d}{down}{bot}{$k_3$}{o1,v1}
       \marrowup{e}{down}{bot}{$k_2$}{i1,v1}
\end{fmfgraph*}
\end{fmffile}
\vspace{10pt}
{\caption{\small The effective four-point interaction. Wavy lines
represent two on-shell gravitons while solid lines represent vector
mesons.}} \label{FigEffVec}
\end{center}
\end{figure}
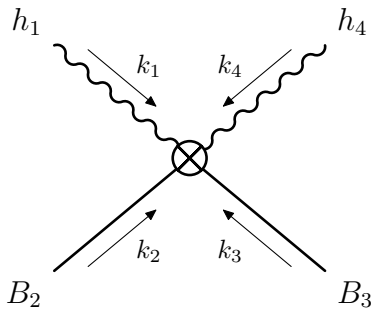
\unitlength=1pt

In the kinematic regime we are interested in, $s \gg t$, the first
order in the scattering amplitude in DIS at small $x$ is given by
terms proportional to $s^2/t$. This is analogous to considering terms
proportional to $s^2$ in the scattering amplitude associated to the
$t$-channel. For dilatons, scalar and vector mesons there is only
one term which has these properties: the one with a factor $(h_1
h_4)\frac{s^2}{t}$. There is another possible term in the vector
case, corresponding to an amplitude proportional to $s^2 [(\e_2 h_1
h_4 e_3) + (\e_3 h_1 h_4 \e_2)]$, but such a term does not appear in
the coefficient of the $t$-channel amplitude. This means that to
this order all terms in the effective Lagrangian above are
sub-leading, and therefore we neglect them. On the other hand, if
we were to consider terms of order $s$, we would have to
include the corresponding terms in the Lagrangian. Then, the only
relevant term within this kinematic regime and at this order is
\beq
\partial_a B_c \partial^b B^c \, \partial^a h^{nq} \partial_b h_{nq}
= \partial_a B_c \partial^b B^c \, \partial^a A^{n} \partial_b A_{n}
\,  v_i v^i \, .
\eeq
%
However, we want to write a gauge invariant Lagrangian. Thus, we use
$G_{ab} = \partial_a B_b - \partial_b B_a$ and $F_{mn} =
\partial_m A_n - \partial_n A_m$. Then, we must consider the following
gauge invariant effective Lagrangian
\beq
G_{ac}G^{bc} \, F^{an}F_{bn} \, , \label{Lvector}
\eeq
which should be multiplied by an appropriate global factor.

\subsection{Hadronic tensor at small $x$ for vector mesons}

Next, we want to explicitly obtain the hadronic tensor for polarized
vector mesons at strong coupling and small $x$ values, and the
corresponding eight structure functions. We will obtain it from
the effective four-point interaction Lagrangian of
Eq.(\ref{Lvector}). By using the same conventions as
\cite{Polchinski:2002jw} for the incoming and outgoing fields we
have to consider the complex conjugate of one $h$ and one $B_a$
fields.

Recall that in the four-point string theory scattering amplitude
${\cal {A}}_4^{2o2c, vector}$ we argue that only the first term is
relevant for the kinematic regime we are interested in. Thus,
effectively we have
\be
{\cal {A}}_4^{2o2c, vector} \simeq  {\cal {P}}_1^{2o2c, vector} \,
{\cal {K}}_1^{2o2c, vector} \, ,
\ee
where ${\cal {K}}_1^{2o2c, vector}$ is given in terms of
Eq.(\ref{Lvector}), while ${\cal {P}}_1^{2o2c, vector}$ can be taken
to be the same as ${\cal {P}}_1^{2o2c, scalar}$ which is the
corresponding pre-factor for scalar mesons. This pre-factor is
associated with the generic dependence of the amplitudes with the
kinematic invariants $s=-u$ and $t$ at small $x$, {\it i.e.}
$t \ll s$ since $x=-\frac{q^2}{2 P \cdot q}$, and also with the
exchange of excited string modes. Thus, we consider
\beq
n_{\mu} n_{\nu} Im_{exc}T^{\mu \nu} = \frac{\pi \alpha'}{8}
\sum_{m=1}^{\infty} \int d\Omega_3 \, dr \, \sqrt{-g} v_{i}v^{i}
G^{*mq}(P)G^{n}_{q}(P) F_{mp}^{*}(q)F_{n}^{p}(q) \delta \left(m-
\frac{\alpha' {\tilde {s}}}{4}\right)\, , \label{iTmunuvector}
\eeq
where all indices are contracted with the full eight-dimensional
metric. Notice that we omit the integrals in the four-dimensional
space $x_0, \cdot \cdot \cdot, x_3$ since they only lead to the
four-momentum conservation delta functions. The solutions to the
field equations are \cite{Koile:2011aa}
\bea
A_{\mu}(q) &=& n_{\mu} f(r) e^{i q\cdot x} \\
A_{r} (q) &=& \frac{-i q\cdot n}{q^2} f'(r) e^{i q\cdot x} \\
B_{\mu}^{l} &=& \frac{\zeta_{\mu}}{\Lambda} \frac{c_i^l}{\Lambda R^3}
\left(\frac{r}{\Lambda R^2}\right)^{-\Delta} Y^l(\Omega_3) e^{i P\cdot x}
= \frac{\zeta_{\mu}}{\Lambda} X^l \\
B_r^{l} &=& 0 \, ,
\eea
where the two first lines are the same as in the scalar field case
since they correspond to the $A^m$ gauge field representing the
virtual photon. As for the scalar mesons we use $\w = q R^2/r$, so
we can can write $f = \w K_1(\w)$, where $K_1$ is the modified
Bessel function of the second kind. We drop
the label $l$ which identifies each meson
\cite{Koile:2011aa,Koile:2013hba}.

The corresponding field strengths are \footnote{Equations for the field
strength $F$ are the same as in the scalar case in Section 2, so we
do not reproduce them here.}
\bea
%
%
%
G_{\mu \nu} &=& i (P_{\mu} B_{\nu} - P_{\nu} B_{\mu}) = \frac{i}{\Lambda} (P_{\mu} \z_{\nu} - P_{\nu} \z_{\mu}) X \\
G_{\mu r} &=& - \partial_{r}B_{\mu} = \frac{\Delta}{\Lambda r}
\z_{\mu} X \, , \,\,\,\,\,\,\,
G_{\mu i} = - \partial_{i}B_{\mu} = \frac{\z_\mu}{\Lambda}
\partial_i X \, ,
\eea
from which we can construct the different pieces involved in the
effective Lagrangian
\bea
G_{rq}^{*}G_{r}^q &=& \frac{R^2}{r^2} \eta^{\rho \sigma}
G_{r\rho}^{*}G_{r \sigma}
= \frac{R^2\Delta^2}{\Lambda^2 r^4} (\z \cdot \z^{*}) X X^{*} \\
G_{\mu q}^{*}G_{r}^q &=& \frac{R^2}{r^2} \eta^{\rho \sigma}
G_{\mu \rho}^{*}G_{r \sigma}
= i \frac{R^2\Delta}{\Lambda^2r^3} (\z \cdot \z^{*}) X X^{*} P_{\mu}\\
G_{\mu q}^{*}G_{\nu}^q &=& \frac{R^2}{r^2} \eta^{\rho \sigma}
G_{\mu \rho}^{*}G_{\nu \sigma} + \frac{r^2}{R^2} G_{\mu r}^{*}G_{\nu r} 
+ \frac{1}{R^2} \tilde{g}^{ij} G_{\mu i} G_{\nu j}\nonumber \\
&=& \frac{R^2}{\Lambda^2r^2} (\z \cdot \z^{*}) X X^{*} P_{\mu}
P_{\nu} + \left(\frac{R^2 P^2}{r^2} + \frac{\Sigma^2}{R^2}\right)
\frac{X X^{*}}{\Lambda^2} \z_{\mu}^{*} \z_{\nu} \, ,
\eea
where we have defined $\Sigma^2 \equiv \Delta^2 + \Delta (\Delta-2)$ due to the presence of the angular derivatives\footnote{Here we have in fact carried out an integration by parts that in this notation should be done later, but the result is the same. We have also used $g_{ij} = R^2\tilde{g}_{ij}$ and some usual spherical harmonic properties.}. Recall that $\Delta$ is a function of the index $l$, and so is $\Sigma$. Next, we have to contract $T^{\mu \nu}$ with the leptonic tensor,
thus all terms with factors $q \cdot n$ and/or $q \cdot n^{*}$ will
vanish. In this way, we only need to expand the effective Lagrangian
$G^{*mq}(P)G^{n}_{q}(P) F_{mp}^{*}(q)F_{n}^{p}(q) = g^{m u} g^{n v}
g^{qq'}g^{pp'} G_{uq}^{*}(P)G_{vq'}(P) F_{mp}^{*}(q)F_{np'}(q)$ in
terms of the solutions of the fields. In the same way as for scalar
mesons we can express the effective four-point interaction
Lagrangian as
\bea
{\cal {L}}_4^{eff,v} &=& {\cal {L}}_{4, A}^{eff,v} + {\cal {L}}_{4,
B}^{eff,v} + {\cal {L}}_{4, C}^{eff,v} + {\cal {L}}_{4, D}^{eff,v}
\, ,
\eea
\bea
{\cal {L}}_{4, A}^{eff,v} &=& \frac{r^4}{R^4} G_{rq}^{*}G_{r}^{q}
F_{rp}^{*}F_{r}^{p}
= \frac{\Delta^2}{\Lambda^2r^2} (f')^2 (\z \cdot \z^{*})(n \cdot n^{*}) |X|^2 \\
{\cal {L}}_{4, B}^{eff,v} &=& {\cal {L}}_{4, C}^{eff,v} = \eta^{\mu
\nu} G_{\mu q}^{*}G_{r}^{q} F_{\mu p}^{*}F_{r}^{p} =\frac{R^4
\Delta}{\Lambda^2 r^5}
f f'  (\z \cdot \z^{*})(n \cdot n^{*}) |X|^2 (P \cdot q)  \\
{\cal {L}}_{4, D}^{eff,v} &=& \frac{R^4}{r^4} \eta^{\mu \rho}
\eta^{\nu \sigma} G_{\rho q}^{*}G_{\sigma}^{q} F_{\mu
p}^{*}F_{\nu}^{p} = \frac{R^4 |X|^2}{r^4 \Lambda^2} \times \left[
\frac{R^4}{r^4} (\z \cdot \z^{*})(n \cdot n^{*}) f^2 (P\cdot q)^2
+ \right. \nonumber \\
&& \left.  \left(\frac{R^4}{r^4} q^2 f^2 + (f')^2\right)(\z \cdot
\z^{*}) (P\cdot n)(P\cdot n^{*}) + \right. \nonumber \\
&& \left. \left( \frac{R^4}{r^4}P^2  + \frac{\Sigma^2}{r^2}\right)
f^2 (n \cdot n^{*}) (q\cdot \z)(q \cdot \z^{*}) + \right. \nonumber \\
&&\left. \left(\frac{R^4}{r^4}P^2q^2f^2 + P^2 (f')^2 +
\frac{\Sigma^2}{r^2}q^2f^2 + \frac{\Sigma^2 r^2}{R^4} (f')^2\right)
(n^{*}\cdot \z^{*}) (n \cdot \z) \right] \, .
\eea
As for the scalar mesons, the terms ${\cal {L}}_{4, A}^{eff,v}$,
${\cal {L}}_{4, B}^{eff,v}$ and ${\cal {L}}_{4, C}^{eff,v}$ are
sub-leading compared with the term in ${\cal {L}}_{4, D}^{eff,v}$
associated with $n \cdot n^{*} $. This is because it is multiplied
by a factor $(P \cdot q)^2$. Therefore, we neglect the contribution
from ${\cal {L}}_{4, A}^{eff,v}$, ${\cal {L}}_{4, B}^{eff,v}$ and
${\cal {L}}_{4, C}^{eff,v}$. The next point is to extract the
polarizations $n$ and $n^{*}$. There are factors in front of the
tensors $\eta^{\mu \nu}$, $P^{\mu}P^{\nu}$ and $\z^{*\mu} \z^{\nu} =
\frac{1}{2} (\z^{*\mu} \z^{\nu} + \z^{*\nu} \z^{\mu}) + \frac{1}{2}
(\z^{*\mu} \z^{\nu} - \z^{*\nu} \z^{\mu})$. The last one contributes
to the anti-symmetric part of the result. Also the condition $(\z^*
\cdot \z)=-P^2$ is satisfied. Then, we arrive to the following
expression
\bea
&& Im_{exc} T^{\mu \nu} = \frac{ \pi \alpha'}{8} \sum_{m=1}^{\infty} \int d\Omega_3 dr 
\sqrt{-g_8} v^i v_i \delta \left(m - \frac{\alpha' \tilde s}{4} \right) \times \nonumber \\
&& P^{\mu}P^{\nu} \times \left[ \frac{R^4 |X|^2}{r^4 \Lambda^2}
\left(\frac{R^4}{r^4} q^2 f^2 + (f')^2\right)(\z \cdot \z^{*}) \right]
+ \left[\frac{1}{2} (\z^{*\mu} \z^{\nu} + \z^{*\nu} \z^{\mu})
+ \frac{1}{2} (\z^{*\mu} \z^{\nu} - \z^{*\nu} \z^{\mu})\right] \nonumber \\
&&  \times\frac{R^4 |X|^2}{r^4 \Lambda^2}\left[\frac{R^4}{r^4}P^2q^2f^2
+ P^2 (f')^2 + \frac{\Sigma^2}{r^2}q^2f^2 + \frac{\Sigma^2 r^2}{R^4}
(f')^2\right] \nonumber \\
&& + \, \eta^{\mu \nu} \times \frac{R^4 |X|^2}{r^4 \Lambda^2}
\left[ \frac{R^4}{r^4}f^2 (\z \cdot \z^{*}) (P\cdot q)^2
+ \left( \frac{R^4}{r^4}P^2 + \frac{\Sigma^2}{r^2}\right)
f^2 (q\cdot \z)(q \cdot \z^{*}) \right. \nonumber \\
&&\left. + (\z \cdot \z^{*}) \left(\frac{r^2 \Delta^2}{R^4} (f')^2 +
2\frac{\Delta}{r} f f' (P\cdot q) \right)\right] \, .
\eea
In order to obtain the structure functions for the
polarized vector mesons we use the Lorentz-tensor decomposition of
the hadronic tensor (\ref{DISWsa}) in terms of its symmetric
and its anti-symmetric parts.

\subsection{New relations of vector meson structure functions}

As for the scalar mesons we use an expression analogous to
(\ref{iTmunu}) for the imaginary part of $T_{\mu\nu}$, which has the
integral in the radial coordinate $r$ and in the angular variables
on a three-sphere. The integrals are:
\bea
I_1&=& \frac{\pi \alpha'}{8} \sum_{m=1}^{\infty} \frac{R^4}{\Lambda^2}
\Sigma^2 q^2 \int d\Omega_3 dr \sqrt{g_8} v^i v_i |X|^2 \frac{f^2(r)}{r^6}
\delta \left(m - \frac{\alpha' s R^2}{4 r^2}\right) \, , \\
I_0&=& \frac{\pi \alpha'}{8} \sum_{m=1}^{\infty} \frac{\Sigma^2}{\Lambda^2}q^2
\int d\Omega_3 dr \sqrt{g_8} v^i v_i |X|^2 \frac{(f'(r))^2}{r^2}
\delta \left(m - \frac{\alpha' s R^2}{4 r^2}\right) \, , \\
\tilde{I}_1 &=& \frac{\pi \alpha'}{8} \sum_{m=1}^{\infty} \frac{R^8}{\Lambda^2}
P^2 q^4 \int d\Omega_3 dr \sqrt{g_8} v^i v_i |X|^2 \frac{f^2(r)}{r^8}
\delta \left(m - \frac{\alpha' s R^2}{4 r^2}\right) \, , \\
\tilde{I}_0 &=& \frac{\pi \alpha'}{8} \sum_{m=1}^{\infty}
\frac{R^4}{\Lambda^2} P^2 q^2 \int d\Omega_3 dr \sqrt{g_8} v^i v_i |X|^2
\frac{(f'(r))^2}{r^4} \delta \left(m - \frac{\alpha' s R^2}{4
r^2}\right) \, .
\eea
The results are as follows
\bea
I_1 &=& \rho_3 |c_i|^2 \frac{\pi \Sigma^2}{8 \Lambda^2 R^2} \left(\frac{\Lambda^2}{q^2}\right)^{\Delta-1}
\frac{1}{\sqrt{4 \pi g_c N}} I_{1,2\Delta+3} \, , \\
I_0 &=& \rho_3 |c_i|^2 \frac{\pi \Sigma^2}{8 \Lambda^2 R^2} \left(\frac{\Lambda^2}{q^2}\right)^{\Delta-1}
\frac{1}{\sqrt{4 \pi g_c N}} I_{0,2\Delta+3} \, , \\
\tilde{I}_1 &=& \rho_3 |c_i|^2 \frac{\pi}{8 \Lambda^2 R^2} \left(\frac{\Lambda^2}{q^2}\right)^{\Delta-1}
\frac{t_B}{\sqrt{4 \pi g_c N}} I_{1,2\Delta+5} \, , \\
\tilde{I}_0 &=& \rho_3 |c_i|^2 \frac{\pi}{8 \Lambda^2 R^2} \left(\frac{\Lambda^2}{q^2}\right)^{\Delta-1} 
\frac{t_B}{\sqrt{4 \pi g_c N}} I_{0,2\Delta+5} \, .
\eea
Since $\tilde{I}_1$ and $\tilde{I}_0$ have a factor $t_B$ they are
sub-leading in comparison with $I_1$ and $I_0$. Then, the structure
functions of polarized vector mesons are
\bea \label{Callan-Gross-initial}
\frac{b_1}{2 \pi} &=& \frac{(P\cdot q)^2}{q^2}[I_1 + \tilde{I}_1] = \frac{1}{4x^2}
[I_1 + \tilde{I}_1] =\frac{I_1}{4x^2} +{\cal {O}}(t) \, ,\\
\frac{b_2}{2 \pi} &=& -(P\cdot q) [I_1 + I_0 + \tilde{I}_1 + \tilde{I}_0]
= \left(1+\frac{I_0}{I_1}\right)\frac{I_1}{2x}+{\cal {O}}(t)  \,  \\
\frac{b_3}{2 \pi} &=& -\frac{2}{3} \frac{b_2}{2 \pi} =- \left(1+\frac{I_0}{I_1}\right)\frac{I_1}{3x}+{\cal {O}}(t) \, , \\
\frac{b_4}{2 \pi} &=& \frac{1}{3} \frac{b_2}{2 \pi} = \left(1+\frac{I_0}{I_1}\right)\frac{I_1}{6x}+{\cal {O}}(t) \, , \\
\frac{g_2}{2 \pi} &=&  \frac{(P\cdot q)^2}{2q^2} [I_1 + I_0 + \tilde{I}_1 + \tilde{I}_0]
= \left(1+\frac{I_0}{I_1}\right)\frac{I_1}{8x^2}+{\cal {O}}(t)   \, \\
\frac{g_1}{2 \pi} &=& -2 \frac{g_2}{2 \pi}=- \left(1+\frac{I_0}{I_1}\right)\frac{I_1}{4x^2}+{\cal {O}}(t)\, ,
\eea
from which we obtain
\bea 
\frac{F_1}{2 \pi} &=& \frac{b_1}{6 \pi}- \frac{P^2}{3 (P \cdot q)}b_2 + \frac{(P \cdot
q)^2}{P^2} \tilde{I}_1
= \frac{I_1}{12x^2}+ {\cal {O}}(t) \, , \\
\frac{F_2}{2 \pi} &=& \frac{b_2}{6 \pi} - \frac{(P\cdot q)q^2}{P^2} [\tilde{I}_1 +
\tilde{I}_0] = \frac{ q^2[I_1 + I_0 + \tilde{I}_1 +
\tilde{I}_0]}{6x} - \frac{q^2 [\tilde{I}_1 + \tilde{I}_0]}{2xP^2}=
\left(1+\frac{I_0}{I_1}\right)\frac{I_1}{6x}+{\cal {O}}(t).\nonumber\\
\eea
Now if we only consider the leading terms, {\it i.e.} we neglect those
proportional to $t_B$, we obtain the following relations
\beq \label{Callan-Gross-Mod}
F_2 = 2x F_1 \left(1+ \frac{I_0}{I_1}\right) \, ,
\eeq
\beq \label{Callan-Gross-Mod-b}
b_2 = 2x b_1 \left(1+ \frac{I_0}{I_1}\right) \, 
\eeq
where since $I_{1, n}= \frac{n+1}{n-1} I_{0, n}$ we have
\beq
1 + \frac{I_1}{I_0} = 1+\frac{I_{1,2\Delta+3}}{I_{0,2\Delta+3}} =
\frac{2\Delta+3}{\Delta+2} \, .
\eeq
Indeed Eq.(\ref{Callan-Gross-Mod}) reproduces the result we obtain
for the scalar mesons. Once more, Eqs.(\ref{Callan-Gross-Mod}) and 
(\ref{Callan-Gross-Mod-b}) are in fact the Callan-Gross relation
with an additional overall factor. Note that
we have obtained a second type of Callan-Gross relation for the
structure functions $b_1$ and $b_2$ given by
Eq.(\ref{Callan-Gross-Mod-b}).

In addition, we also obtain the relations:
\beq
b_1= 3 F_1 \, , \hspace{1.cm} b_4= -\frac{1}{2}
b_3 \, , \hspace{1.cm} g_1= -2 g_2 \, .
\eeq

On the other hand, in the regime $x\ll \exp{(-\sqrt{\lambda})}$ the
calculations are similar to those presented in section 2.6.

\section{General results for different Dp-brane models}

In this section we generalize the results of the previous sections. 
Thus, the results presented here hold for
both type IIA and type IIB string theory dual models with one flavor
Dp-brane, corresponding to the D3D7-brane \cite{Kruczenski:2003be},
D4D8$\overline{D8}$-brane, \cite{Sakai:2004cn},
D4D6$\overline{D6}$-brane models \cite{Kruczenski:2003uq}.
We consider the generic asymptotic induced metric
\beq
ds^2 = \left(\frac{r}{R}\right)^{\alpha} \eta_{\mu \nu} dx^{\mu}
dx^{\nu} + \left(\frac{r}{R}\right)^{\beta} \left[dr^2 + r^2
d\Omega_{p-4}^2\right] \, ,
\eeq
where the parameters depend on the specific  model as listed
below\footnote{This generic form of the metric holds only for $r\gg
L$, being $L$ some dimensionful parameter in each background. We
can do this because in the local approximation the interaction in which we are interested
occurs for large values of $r$.}
\begin{center}
\centering
\begin{tabular}{|c|c|c|c|}
\hline
Model $/$ Parameter & p & $\alpha$ & $\beta$ \\
\hline
$D3D7$ & 7 & 2 & -2 \\
\hline
$D4D8\overline{D8}$ & 8 & $3/2$ & $-3/2$ \\
\hline
$D4D6\overline{D6}$ & 6 & $3/2$ & $-3/2$ \\
\hline
\end{tabular}
\end{center}
the corresponding fields were obtained in our previous paper
\cite{Koile:2013hba}
\bea
X^{l} &=& \frac{c_i}{\Lambda R^{\frac{p-1}{2}}} e^{iP\cdot x} \left(\frac{r}{R\Lambda^2}\right)^{A-\gamma B}  Y^{l}(S^{p-4}), \\
B_{\mu}^{l} &=& \frac{\z_{\mu}}{\Lambda} X^{l}, \\
A_{\mu} &=& n_{\mu} e^{i q\cdot y} f(r), \\
A_{r} &=& \frac{-i}{q^2}(q \cdot n) e^{i q\cdot y} f'(r), \\
f(r) &=& \Gamma^{-1}(n+1)
\left(\frac{qR}{2B}\right)^{n+1}\left(\frac{r}{R}\right)^{-B(n+1)}K_{n+1}\left[
\frac{qR}{B} \left(\frac{r}{R}\right)^{-B}\right] \, ,
\eea
where  $A = \frac{1}{2}(1-\theta)$, $\theta = 2 \alpha +
(p-3)\beta/2 + (p-4)$, $\gamma^2 = \frac{A^2 + l (l+p-5)}{B^2}$, $B=
\frac{1}{2}(\alpha-\beta-2)$ and $n=\frac{2+\beta}{4B}$. We can make
the notation simpler by defining $\w \equiv \frac{qR}{B}
\left(\frac{R}{r}\right)^{B}$, therefore $f(r) = \Gamma^{-1}(n+1)
\w^{n+1} K_{n+1}(\w)$. Notice that in the notation of the previous sections
we have used $\Delta = \gamma B - A$.
As in the previous section the starting point is
\beq \label{tensorT}
n_{\mu} n_{\nu} Im_{exc}T^{\mu \nu} = \frac{\pi \alpha'}{8}
\sum_{m=1}^{\infty} \int d\Omega_{p-4} dr \sqrt{-g_{p+1}} v_{i} v^{i}
G^{*ac}(P)G^{b}_{c}(P) F_{ap}^{*}(q)F_{b}^{p}(q) \delta\left(m-\frac{\alpha' \tilde s}{4}\right) \, ,
\eeq
and we omit the Dirac delta functions associated with the momentum
conservation since we have already taken them into account whenever we
set the momenta of the external particles.

In the general case the Bessel functions satisfy the identity
\beq
\partial_{\w}(\w^{n+1} K_{n+1} (\w)) =  (n+1) \w^{n} K_{n+1}(\w)
- \w^{n+1}\left( \frac{n+1}{\w}K_{n+1}(\w) + K_n(\w) \right) = - \w^{n+1} K_{n}(\w),
\eeq
for the derivative of $f(r)$ we obtain
\beq
f'(r) = \frac{1}{\Gamma(n+1)}\frac{d}{d\w}(\w^{n+1} K_{n+1} (\w))
\frac{d\w}{dr}  = \frac{1}{\Gamma(n+1)} \frac{B}{R}
\left(\frac{B\w}{qR} \right)^{1/B} \w^{n+2} K_n (\w) \, .
\eeq
Now we are in conditions to obtain the structure functions in this
small $x$ region. We only need to know the integral\footnote{Recall that this result 
is valid provided that the arguments of the gamma functions are positive.}
\bea
\int_0^\infty d\w \, \w^D \, K^2_n(\w) =  I_{n,D} &=& 2^{D-2}
\frac{\Gamma(\nu+n)\Gamma(\nu-n)\Gamma^2(\nu)}{\Gamma(2\nu)} \ , \
\nu = \frac{1}{2}(D+1),
\eea
By defining integrals analogous to Eqs.(\ref{I1a}) and (\ref{I2a}) 
\bea
I_1&=& \frac{\pi \alpha'}{8} \sum_{m=1}^{\infty} R^6
q^2 \int d\Omega_{p-4} dr \sqrt{g_{p+1}} v_i v^i|X|^2 \frac{f^2(r)}{r^6}
\delta \left(m - \frac{\alpha' s R^2}{4 r^2}\right) \, , \\
I_0&=& \frac{\pi \alpha'}{8} \sum_{m=1}^{\infty} R^2 \int d\Omega_{p-4}
dr \sqrt{g_{p+1}} v_i v^i|X|^2 \frac{(f'(r))^2}{r^2}
\delta \left(m - \frac{\alpha' s R^2}{4 r^2}\right) \, .
\eea
we obtain
\bea
&&I_1\propto I_{n+1,D} \ , \ I_0  \propto I_{n,D}\\
D  &=&  2n+1 + \frac{1}{B}\left[ 2 \Delta + \alpha - (\beta/2 + 1)(p-5)\right] , \nonumber\\
&&\tilde{I}_1  \propto  I_{n+1,\tilde{D}}  \ , \ \tilde{I}_0  \propto  I_{n,\tilde{D}}  \\
\tilde{D}  &=&  2n+3 + \frac{1}{B}\left[ 2 \Delta + 2\alpha -
(\beta/2 + 1)(p-3)\right] , \nonumber
\eea
If we consider the case of the D3D7-brane system ($\beta = -2$) the last term disappears. On the other hand,
for the D4D8$\mathrm{\overline{D8}}$- and D4D6$\mathrm{\overline{D6}}$-brane models we simply have to set $\alpha = 3/2 = - \beta$, and
$n=1/4$. 

All these integrals are
convergent, therefore the $x$-dependence on the structure functions is the
same as before and coincides for all models that we study. Thus, in a sense we
can consider a sort of universal behaviour of the relations among the
structure functions for scalar and polarized vector mesons. The only
thing which changes for these relations is the factor
$(1+\frac{I_1}{I_0})$. The two relevant integrals are
\bea
I_1 &=& \frac{\pi \alpha^2 \Sigma^2 \rho_{p-4} |c_i|^2 q^4 }{8
\sqrt{4 \pi g_c N}\Lambda^4 \Gamma^2 (n+1) B}   \left(\frac{qR}{B}\right)^{(2n+1)-D} I_{n+1,D} \, , \\
I_0 &=& \frac{\pi \alpha^2 B \Sigma^2 \rho_{p-4} |c_i|^2 q^2}{8
\sqrt{4 \pi g_c N} \Lambda^4R^2 \Gamma^2 (n+1)}  
\left(\frac{qR}{B}\right)^{(2n+3)-D} I_{n,D} \, ,
\eea
where we have defined $\Sigma^2\equiv\Delta^2+l(l+p-5)$.

From these expressions the results for the structure functions are
the same as in the previous section, just replacing the new values
for $I_1$ and $I_0$. For the Callan-Gross like relation we can
calculate the generic factor
\beq
1 + \frac{I_0}{I_1} = 1 + \frac{B^2}{q^2R^2}
\left(\frac{qR}{B}\right)^{2} \frac{I_{n,D}}{I_{n+1,D}} =
\frac{2D}{D+2n+1} \, ,
\eeq
which leads to a function which takes values between one and two.
In addition, the general scalar meson case is easy to derive.
Also, notice that by setting $D=2\Delta+3$ and $n=0$ we recover the
expression obtained in the $D3D7$-brane model. On the other hand,
the factors $\frac{16(\Delta+1)}{8\Delta+11}$ and $\frac{4(4\Delta +
3)}{8 \Delta + 9}$ are obtained for  $D4D6\overline{D6}$-brane
$D4D8\overline{D8}$-brane models, respectively.

\newpage

\section{Conclusions and discussion}

In this work we have investigated deep inelastic scattering of leptons 
from spin-zero and spin-one hadrons at small
values of the Bjorken parameter $x$ in terms of superstring theory, 
by using the gauge/string duality. We have considered single-flavored 
scalar and vector mesons in the large $N$ limit and at the strongly coupled
regime of the gauge field theories. This has been studied for different
holographic dual models with flavor Dp-branes in type IIA and type
IIB superstring theories. We have derived the hadronic tensor
and the structure functions for scalar and polarized vector mesons,
the latter having a very rich Lorentz-tensor structure.
For polarized spin-one mesons we have obtained the eight structure functions
at small values of the Bjorken parameter. Our most interesting result is that we
found new relations of the Callan-Gross type for several structure
functions. These relations have similarities for all different
Dp-brane models that we have investigated. We think that this can be
interpreted as a signal of a universal behavior for confining gauge theories
which have a string theory dual description where mesons can be engineered 
with flavor Dp-branes in the probe approximation. It is worth emphasizing that
this effect does not depend on supersymmetry nor on conformal symmetry of the
dual gauge field theories. It can be related to the isometries of the background:
recall that in order to calculate the tree-level Feynman diagrams for the
$t$-, $s$-, $u$-channels and the contact term we consider the interaction
Lagrangian derived from the DBI-action of the probe flavor Dp-brane. This Dp-brane
wraps a cycle, $S^{p-4}$. In the Kaluza-Klein decomposition of the fields
defined on the Dp-brane there are spherical harmonics factors $Y^l(\Omega_{p-4})$, which
satisfy the eingenvalue equations 
$i Q_l Y^l(\Omega_{p-4}) = v^i \partial_i Y^l(\Omega_{p-4})$.  
From this one can construct the graviton-meson-meson interaction, by using the
N\"other's theorem. This in fact is what we have done in the parametric regime
where $1/\sqrt{\lambda} \ll x < 1$ \cite{Koile:2011aa,Koile:2013hba}.

As for the glueballs \cite{Polchinski:2002jw} we have found 
four kinematical regimes in terms of the 
Bjorken parameter. When $1/\sqrt\lambda \ll x < 1$ the holographic 
dual description is in terms of supergravity and we have studied it
in our previous papers \cite{Koile:2011aa,Koile:2013hba}.
For this regime we found several relations among the structure functions.
We obtained $2 F_1(x)=F_2(x)$ and $2 b_1(x)=b_2(x)$, 
which we conjectured to be the pure supergravity 
version of $2 x F_1(x) \sim F_2(x)$ and $2 x b_1(x) \sim b_2(x)$. 
In the present work, for small $x$, 
we have found that in addition to the Bjorken parameter factor there 
is another factor which depends on the scaling dimensions $\Delta$ of the 
meson wavefunctions, which does not dependent on $x$ as shown
in Eqs.(\ref{Callan-Gross-Mod}) and (\ref{Callan-Gross-Mod-b}), respectively.
Also, in \cite{Koile:2011aa}
we obtained the relations: $2 b_3 = -b_4$, $2 g_2 = -
g_1$, $b_2=3 F_2$ and $b_1 =3 F_1$, for $1/\sqrt\lambda \ll x < 1$ 
in the $t_B \rightarrow 0$ limit. Now, for small and exponentially small
$x$ some of these last relations are slightly different as shown
in Section 3 of the present work. Notice that for small $x$ we have obtained
Eqs.(\ref{Callan-Gross-initial}). By comparison we observe that the relation
between $g_1$ and $g_2$ is preserved while the rest of relations become
modified as we reduce the values of the Bjorken parameter towards small and
exponentially small values.
So, in the second kinematic regime corresponding to $\exp{(-\sqrt\lambda)} \ll x \ll
1/\sqrt\lambda$ the holographic dual description goes beyond
the supergravity approximation and it includes excited strings. 
The third regime $x \ll \exp{(-\sqrt\lambda)}$ corresponds to
the case when the size of the strings is comparable to the scale of the 
anti-de Sitter space. Within this regime the interaction becomes non-local
and the strings growth is taken into account by a diffusion operator.
There is a fourth regime where $|\ln x| \lambda^{-1/2} > \ln (\Lambda/q)$, 
where $q$ is the momentum transfer and $\Lambda$ is the confining IR scale. 
Here is where the world-sheet renormalization group can be used to include
the effect of strings growth. The general picture in the planar limit 
of the strongly coupled gauge field theory corresponds to the scattering 
of a lepton by an entire hadron. We would expect that deeper understanding
of string theory description in the exponentially small regime will 
reveal the parton structure, and therefore we should recover the growth of
the structure functions with $q^2$. We should keep in mind that the string
theory scattering amplitudes we consider are in flat ten-dimensional spacetime,
and that as in \cite{Polchinski:2002jw} we fold it into the AdS wavefunctions.
The idea is that the momentum invariants in the inertial frame are of order
the string scale, therefore, it is expected that the scattering process
should be localized on this scale. This scale is small in comparison
with the AdS scale $R$, so one can take the string theory
scattering amplitude in flat space and then consider the warped wavefunctions
of the fields. On the other hand, for the exponentially small $x$ region this
approximation is not longer valid and a diffusion operator plays an important role.

We have answered several questions. On the one hand, we have obtained
the eight structure functions from holographic dynamical hadrons, and
investigated the $x$-dependence of the structure functions
for small and exponentially small values of the Bjorken parameter,
in the planar limit and at strong coupling of the gauge theories.
The $q^2$-dependence we have obtained for $0 < x \ll
1/\sqrt{\lambda}$ for both scalar and vector mesons structure
functions is the one expected from the operator product expansion in
the large $N$ limit at strong coupling. 
As pointed out for glueballs \cite{Polchinski:2002jw} by considering
string scattering on flat spacetime instead of a power-low falling
with $q^2$ one obtains an exponential falling. This indicates a very
soft amplitude, and the reason is that there are no partons in this
case. On the other hand, in our calculations, similarly to what
happens for the glueballs the key point is that the curved geometry
induces a power law behavior. Basically, as the string goes from the
bulk to the boundary its tension increases, and consequently its
size becomes smaller, of order of the inverse momentum transfer.

We must emphasize that the regime in the dual gauge theory
that we are exploring is the strongly coupled regime $N \gg \lambda
\gg 1$, and moreover this is studied in the planar limit. Then,
hadronic structure functions are calculated from a forward
scattering amplitude, and therefore we must consider the Regge
physics of the string. In terms of the bulk theory, when $x \sim
1/\sqrt{\lambda}$ it turns out that for a local observer in the bulk
the energy scale becomes the string mass scale since $\tilde s \sim
1/\alpha'$. So, in this parametric region the string dynamics in the
bulk becomes very important. Also notice that in QCD the Pomeron
exchange dominates the gluon structure function for the regime $s
\approx q^2/x$ for small $x$. The Pomeron is a
conjectured trajectory of glueball states. In string theory
something similar happens with the gravitons, where now one has to
consider the lowest mode of the graviton in the curved space-time,
which has an IR cutoff. This IR cutoff leads to a mass gap of order
$1/R^2$ and therefore the intercept becomes $\alpha'/R^2$ which is
order $1/\sqrt{\lambda}$. The contribution of the Pomeron to the
parton distribution functions, which is proportional to
$x^{-\alpha_0}$, is given by the intercept of the Pomeron trajectory
$\alpha_0$. Then, it is expected to have the behavior $F_1 \propto
x^{-2+{\cal {O}}(1/\sqrt{\lambda})}$ and $F_2 \propto x^{-1+{\cal
{O}}(1/\sqrt{\lambda})}$ that we have found. 

There are many open questions. One is to understand better the 
approximation we have made in order to keep only the
first term in the four-point string theory scattering amplitude, 
particularly for spin-one hadrons. We expect to investigate it
further in terms of the operator product expansion of the string
theory vertex operators on the string worldsheet
\cite{Brower:2006ea,Cornalba:2008sp,Cornalba:2009ax,Cornalba:2010vk}.
Another issue we plan to report in a forthcoming work is
about the implications on observables such as the 
scattering cross sections for DIS \cite{koile3}.

~

~

\centerline{\large{\bf Acknowledgments}}

~

We thank Sebasti\'an Macaluso for collaboration in an early stage of
this project. We thank Carlos N\'u\~nez and Sebasti\'an Macaluso for
a critical reading of the manuscript and useful comments. 
This work has been partially supported by the
CONICET-PIP 0595/13 Grant.


\newpage


\begin{thebibliography}{99}




\bibitem{Polchinski:2002jw}
  J.~Polchinski, M.~J.~Strassler,
  ``Deep inelastic scattering and gauge / string duality,''
  JHEP {\bf 0305 } (2003)  012.
  [hep-th/0209211].

\bibitem{Manohar:1992tz}
  A.~V.~Manohar,
  ``An Introduction to spin dependent deep inelastic scattering,''
  [hep-ph/9204208].

\bibitem{Hoodbhoy:1988am}
  P.~Hoodbhoy, R.~L.~Jaffe, A.~Manohar,
  ``Novel Effects in Deep Inelastic Scattering from Spin 1 Hadrons,''
  Nucl.\ Phys.\  {\bf B312 } (1989)  571.

\bibitem{Hatta:2007cs}
  Y.~Hatta, E.~Iancu, A.~H.~Mueller,
  ``Deep inelastic scattering off a N=4 SYM plasma at strong coupling,''
  JHEP {\bf 0801 } (2008)  063.
  [arXiv:0710.5297 [hep-th]].

\bibitem{Hassanain:2009xw}
  B.~Hassanain, M.~Schvellinger,
  ``Holographic current correlators at finite coupling and scattering off a supersymmetric plasma,''
  JHEP {\bf 1004 } (2010)  012.
  [arXiv:0912.4704 [hep-th]].
  
\bibitem{CaronHuot:2006te}
  S.~Caron-Huot, P.~Kovtun, G.~D.~Moore, A.~Starinets and L.~G.~Yaffe,
  ``Photon and dilepton production in supersymmetric Yang-Mills plasma,''
  JHEP {\bf 0612} (2006) 015
  [arXiv:hep-th/0607237].

\bibitem{Hassanain:2011ce}
  B.~Hassanain and M.~Schvellinger,
  ``Diagnostics of plasma photoemission at strong coupling,''
  Phys.\ Rev.\ D {\bf 85} (2012) 086007
  [arXiv:1110.0526 [hep-th]].

\bibitem{Hassanain:2011fn}
  B.~Hassanain and M.~Schvellinger,
  ``Plasma conductivity at finite coupling,''
  JHEP {\bf 1201} (2012) 114
  [arXiv:1108.6306 [hep-th]].

\bibitem{Hassanain:2010fv}
  B.~Hassanain, M.~Schvellinger,
  ``Towards 't Hooft parameter corrections to charge transport in strongly-coupled plasma,''
  JHEP {\bf 1010 } (2010)  068
  [arXiv:1006.5480 [hep-th]].

\bibitem{Hassanain:2012uj}
  B.~Hassanain and M.~Schvellinger,
  ``Plasma photoemission from string theory,''
  JHEP {\bf 1212} (2012) 095
  [arXiv:1209.0427 [hep-th]].

\bibitem{Polchinski:2000uf}
  J.~Polchinski and M.~J.~Strassler,
  ``The String dual of a confining four-dimensional gauge theory,''
  hep-th/0003136.

\bibitem{Kogut:1974ni}
  J.~B.~Kogut and L.~Susskind,
  ``Scale invariant parton model,''
  Phys.\ Rev.\ D {\bf 9} (1974) 697.

\bibitem{Koile:2011aa}
  E.~Koile, S.~Macaluso and M.~Schvellinger,
  ``Deep Inelastic Scattering from Holographic Spin-One Hadrons,''
  JHEP {\bf 1202} (2012) 103
  [arXiv:1112.1459 [hep-th]].

\bibitem{Koile:2013hba}
  E.~Koile, S.~Macaluso and M.~Schvellinger,
  ``Deep inelastic scattering structure functions of holographic spin-1 hadrons with $N_f \geq 1$,''
  JHEP {\bf 1401} (2014) 166
  [arXiv:1311.2601 [hep-th]].

\bibitem{Kruczenski:2003be}
  M.~Kruczenski, D.~Mateos, R.~C.~Myers, D.~J.~Winters,
  ``Meson spectroscopy in AdS / CFT with flavor,''
  JHEP {\bf 0307 } (2003)  049.
  [arXiv:hep-th/0304032 [hep-th]].

\bibitem{Sakai:2004cn}
  T.~Sakai, S.~Sugimoto,
  ``Low energy hadron physics in holographic QCD,''
  Prog.\ Theor.\ Phys.\  {\bf 113 } (2005)  843-882.
  [arXiv:hep-th/0412141 [hep-th]].

\bibitem{Kruczenski:2003uq}
  M.~Kruczenski, D.~Mateos, R.~C.~Myers and D.~J.~Winters,
  ``Towards a holographic dual of large N(c) QCD,''
  JHEP {\bf 0405} (2004) 041
  [hep-th/0311270].

\bibitem{Stieberger:2009hq}
  S.~Stieberger,
  ``Open \& Closed vs. Pure Open String Disk Amplitudes,''
  arXiv:0907.2211 [hep-th].

\bibitem{Hashimoto:1996kf}
  A.~Hashimoto and I.~R.~Klebanov,
  ``Decay of excited D-branes,''
  Phys.\ Lett.\ B {\bf 381} (1996) 437
  [hep-th/9604065].

\bibitem{Fotopoulos:2002wy}
  A.~Fotopoulos and A.~A.~Tseytlin,
  ``On gravitational couplings in D-brane action,''
  JHEP {\bf 0212} (2002) 001
  [hep-th/0211101].

\bibitem{Brower:2006ea}
  R.~C.~Brower, J.~Polchinski, M.~J.~Strassler and C.~I.~Tan,
  ``The Pomeron and gauge/string duality,''
  JHEP {\bf 0712} (2007) 005
  [hep-th/0603115].

\bibitem{Cheung:2010vn} 
  C.~Cheung, D.~O'Connell and B.~Wecht,
  ``BCFW Recursion Relations and String Theory,''
  JHEP {\bf 1009}, 052 (2010)
  [arXiv:1002.4674 [hep-th]].

\bibitem{Fotopoulos:2010cm} 
  A.~Fotopoulos and N.~Prezas,
  ``Pomerons and BCFW recursion relations for strings on D-branes,''
  Nucl.\ Phys.\ B {\bf 845}, 340 (2011)
  [arXiv:1009.3903 [hep-th]].

\bibitem{Gross:1986mw}
  D.~J.~Gross and J.~H.~Sloan,
  ``The Quartic Effective Action for the Heterotic String,''
  Nucl.\ Phys.\ B {\bf 291} (1987) 41.

\bibitem{Sannan:1986tz}
  S.~Sannan,
  ``Gravity as the Limit of the Type {II} Superstring Theory,''
  Phys.\ Rev.\ D {\bf 34} (1986) 1749.

\bibitem{Green:1987sp}
  M.~B.~Green, J.~H.~Schwarz and E.~Witten,
  ``Superstring Theory. Vol. 1: Introduction,''

\bibitem{Peeters:2006kp}
  K.~Peeters,
  ``A Field-theory motivated approach to symbolic computer algebra,''
  Comput.\ Phys.\ Commun.\  {\bf 176} (2007) 550
  [cs/0608005 [cs.SC]].

\bibitem{Peeters:2007wn}
  K.~Peeters,
  ``Introducing Cadabra: A Symbolic computer algebra system for field theory problems,''
  hep-th/0701238 [HEP-TH].



\bibitem{Cornalba:2008sp}
  L.~Cornalba and M.~S.~Costa,
  ``Saturation in Deep Inelastic Scattering from AdS/CFT,''
  Phys.\ Rev.\ D {\bf 78} (2008) 096010
  [arXiv:0804.1562 [hep-ph]].

\bibitem{Cornalba:2009ax}
  L.~Cornalba, M.~S.~Costa and J.~Penedones,
  ``Deep Inelastic Scattering in Conformal QCD,''
  JHEP {\bf 1003} (2010) 133
  [arXiv:0911.0043 [hep-th]].

\bibitem{Cornalba:2010vk}
  L.~Cornalba, M.~S.~Costa and J.~Penedones,
  ``AdS black disk model for small-x DIS,''
  Phys.\ Rev.\ Lett.\  {\bf 105} (2010) 072003
  [arXiv:1001.1157 [hep-ph]].

\bibitem{koile3}
E.~Koile, N.~Kovensky and M.~Schvellinger, Work in preparation.


\end{thebibliography}
\end{document}